***Studying Complexity in Solar Wind Plasma***

***During Shock Events.***

***Part I: Nonextensive Tsallis Statistics***

G.P. Pavlos[1], A.C. Iliopoulos[1], G.N. Zastenker[2] , L.M. Zelenyi[2], L.P. Karakatsanis[1], M. Riazantseva[2], M.N. Xenakis[3], E.G. Pavlos[4]

(1) Democritus University of Thrace, Department of Electrical Engineering, Xanthi
(2) Space Research Institute (IKI)
(3) German Research School for Simulation Sciences, Aachen, Germany
(4) Department of Physics, Aristotle University of Thessaloniki, 54624 Thessaloniki, Greece

Corresponding author: A.C. Iliopoulos, DUTH, (ailiopou@gmail.com)

## ABSTRACT

Novel results which reveal phase transition processes in the solar wind plasma during shock events are presented in this study which is the first part of a trilogy concerning the solar wind complexity. Solar wind plasma is a typical case of stochastic spatiotemporal distribution of physical magnitudes such as force fields ($\mathbf{B}$, $\mathbf{E}$) and matter fields (particle and current densities or bulk plasma distributions). The results of this study can be understood in the framework of modern theoretical concepts such as non-extensive statistical mechanics (Tsallis, 2009), fractal topology (Zelenyi and Milovanov, 2004), turbulence theory (Frisch,1996), strange dynamics (Zaslavsky, 2002), percolation theory (Milovanov, 1997), anomalous diffusion theory and anomalous transport theory (Milovanov, 2001), fractional dynamics (Tarasov, 2007) and non-equilibrium phase transition theory (Chang, 1992). This study shows clearly the non-extensive and non-Gaussian character of the solar wind plasma and the existence of multi-scale strong correlations from the microscopic to the macroscopic level. This result indicates the inefficiency of classical MHD or plasma statistical theories based on the classical central limit theorem to explain the complexity of the solar wind dynamics, since these theories include smooth and differentiable spatial-temporal functions (MHD theory) or Gaussian statistics (Boltzmann-Maxwell statistical mechanics). However, the results of this study indicate the presence of non-Gaussian non-extensive statistics with heavy tails probability distribution functions, which are related to the *q*-extension of central limit theorem.

## 1. Introduction

This is the first part of a trilogy of papers concerning the solar wind complexity during shock events. Solar Wind includes ionized and magnetized gas, composed mainly by protons, electrons, alpha particles and heavier ions, continuously flowing away from the solar corona in all directions pervading the interplanetary space. Parker (Parker, 1958) conjectured that hot solar corona undergoes a steady expansion, while the low value plasma velocity near the Sun increases to a large supersonic expansion speed at large heliocentric distances. Because of the high electrical conductivity of the coronal plasma the magnetic field lines are frozen in the bulk plasma flow, which transports the magnetic field lines into the interplanetary space. Spacecraft measurements showed the presence of low frequency and large amplitude magnetic field fluctuations obeying power laws in the energy spectrum indicating that turbulence at the MHD scales is a characteristic feature of the solar wind system. This feature implies that there are mechanisms for energy loss from large scale plasma structures through cascade processes at intermediate scales and exciting kinetic processes at small scales. Consequently the MHD turbulence may play essential role in heating and acceleration processes in the solar wind. After all turbulence of the solar wind is the manifestation, in a deeper level, of the complexity and strange dynamics of the solar wind plasma system.

Concerning MHD turbulence and complex phenomena in Solar Wind (SW) dynamical system, excellent review articles were produced by Marsch and Tu (March, 1997), Bruno and Carbone (Bruno, 2005), Carbone Veltri and Bruno (Carbone, 1996), Zelenyi and Milovanov,( Zelenyi, 2004). In particular, according to Milovanov and Zelenyi (Milovanov,1994a;1994b), Zelenyi and Milovanov (Zelenyi,1991;2004) the complexity of solar wind, as it was described by Burlaga (Burlaga,1986;1984), Burlaga and Klein (Burlaga,1991), Tu and Marsch (Tu, 1995) Marsch and Tu (March,1997) and others, can de related to the fractal character  of the solar photosphere and convection zone.

Traditionally the quantities of key interest in turbulence theory are two point correlation functions and structure functions of the vector or scalar fields. These quantities depend on the field differences between two points in space and time, allowing resolving a broad range of spatio-temporal scales of the turbulent fluctuations. According to Marsch and Tu (1997), the fluctuating fields in the solar wind are often large in the moving frame and from a practical point of view their time histories resemble strongly the time evolution of mathematical random variables. The observed small-scale fluctuations of solar wind MHD quantities appear to be composed of waves and connecting pressure-balanced structures which occur as short-lived and nonlinear coupled excitations on a broad range of scales. Temperature fluctuations were detected and found to play a prominent role in the local pressure balance, often distinctly anti-correlated with the magnetic field amplitude fluctuations and well correlated with the solar wind speed. This signature may stem from coronal heating and pressure-equilibration processes to adjacent plasma flow tubes near the Sun. The turbulent are scale dependent and vary in close connection with the interplanetary magnetic field and the stream structure.

Solar wind turbulent state can be explained through the Tsallis non-extensive generalization of statistical mechanics and the Tsallis q-entropy principle underlying the self-organization process of space plasmas.

Complexity theory is referred to the ordering principle that acts in a nonlinear system causing its self organization. This principle is manifested through the entropy principle which drives the dynamics of the system to equilibrium or meta-equilibrium states of maximum entropy. Tsallis non-extensive statistical mechanics (Tsallis 1988, 2004b) includes the generalization of Boltzmann-Gibbs statistical mechanics, based at the multifractal theory for extending Boltzmann-Gibbs entropy to the Tsallis $q$-entropy where the B-G statistical theory is a special case corresponding to the value $q$=1. The Tsallis $q$-entropy is related to strange kinetics and fractal-multifractal profile of the phase space. This causes long range spatio-temporal correlations and non-Gaussian probability distribution functions of the dynamical fluctuations as well as non-Poissonian temporal distributions (Milovanov *et al.* 1993; Zaslavsky, 2002). These characteristics, also with critical exponents, power laws or heavy tails of probability distribution functions, as well as singularities and spatial-temporal fractality of non-differentiable distribution of the physical functions and physical properties-quantities are related to the strange topology of phase space caused by the $q$-entropy principles of Tsallis theory as the dynamical system tries to achieve extremization of $q$-entropy states (Alemany & Zanette, 1994).

The fractal topology of the phase space is produced by the fractional dynamics of the system which is related to anomalous diffusion processes, multiscale cooperation and correlations as well as development of memory processes and percolation clusters (Zelenyi and Milovanov, 2004; Milovanov, 2012). The fractional dynamics in phase space is related or produced also by non-ergodic and intermittent self-similar hierarchies of islands, cantories, stirrers and traps, which can create spatial-temporal coherence and intermittency, through Levy type processes (Zaslavsky 2002).

The intermittent turbulent states and the spatial-temporal behavior of nonlinear distributed dynamical systems, such as the solar wind system, can be considered as the spatial-temporal mirroring of the strange or fractal topology of the system dynamical phase space and the included strange fractional dynamics in the phase space. As the control parameters of the system change, the system can generate topological and dynamical phase transition processes developing to new equilibrium or meta-equilibrium states corresponding to local extremization of $q$-entropy and new fixed points, according to non-equilibrium renormalization theory (Chang *et al.*, 2003).

The last decades many scientists have worked for the understanding of the solar wind plasma dynamics through the complexity theory. The concept of self-organization and low-dimensional chaotic process for solar wind and space plasmas was supported by Burlaga in a series of novel studies (Burlaga 1987, 1990a,b, 1991a,b,c, 1992, 1993, Burlaga and Forman 2002, Burlaga et al. 2003) and Pavlos et al. (1991; 1992a,b). Moreover Pavlos et al. (Pavlos, 2011;2012) presents the multiscale and multifractal and non-Gaussian character of the solar wind medium introducing the intermittent turbulence theory for the explanation of solar wind dynamics.

Burlaga (1991a,b,c,1992) showed that speed fluctuations bear a multifractal structure in recurrent streams measured between 1 and 6 AU. The large scale magnetic field fluctuations as observed in the outer heliosphere (~25 AU) were found to have a multifractal formation (Burlaga 1991b). Also, small-scale velocity fluctuations were observed near ~8.5 AU including multifractal structure, suggesting that solar wind turbulence consist of a mixture of sheets and space-filling eddies of various sizes. The concept of intermittent turbulence process in the solar wind plasma was further studied by many scientists (Marsch and Liu, 1993; Ruzmakin *et al.*, 1995; Tu *et al.*, 1996; Carbone *et al.*, 1996; Horbury & Balogh, 1997; Strunik and Macek, 2008; Riazantseva *et al*. 2010 and references within).

Theoretical models for explaining the self similar, multiscale, multifractal and intermittent MHD turbulent character of Solar Wind system were also developed by many scientists (Carbone, 1993, 1994; Buti, 1996; Carbone *et al.*, 1996; Marsh and Tu, 1997; Macek, 2006, 2007; Macek *et al*, 2012; Sorriso Valvo *et al.*, 1999, 2007; Leitner *et al*. 2009).

Milovanov and Zelenyi (1994a,b, 1999) and Zelenyi and Milovanov (2004) introduced fractional models of the self-affine and fractal temporal and spatial distribution of solar wind random magnitudes. In addition, Zelenyi and Milovanov (1991) and Milovanov and Zelenyi (Milovanov, 1993), through the introduction of fracton excitations, managed to discriminate between the internal fractional processes of the solar wind as a driving mechanism of the self-organized solar wind dynamics and the fractional temporal and spatial characteristics caused by the photospheric fractional dynamics.

More generally, according to Zelenyi and Milovanov (2004) the complex character of the solar wind system includes the existence of nonequilibrium (quasi)-stationary states (NESS) having the topology of a percolating fractal set. The stabilization of a system near the NESS is perceived as a transition into a turbulent state determined by self-organization processes. The large-scale order of the NESS turbulent state can be identified with the generalized symmetries of a fractal disk diffeomorphic to a fractal set at the percolation threshold (Milovanov, Phys.Rev.E56, 2437, 1997). The structural stability of the NESS, as a symmetric turbulent phase, is maintained due to multi-scale correlations with divergence of the correlation length for the fractal distribution $\xi \rightarrow \infty$. The long-range correlation effects manifest themselves as a strange non-Gaussian behavior of kinetic processes near the NESS plasma state.

Moreover, the non-Gaussian character of solar wind fluctuations can be explained using the nonextensive statistical theoretical framework introduced by Tsallis (Tsallis, 1988; 2004b). Pavlos et al (2013…teleutaio) have shown the non-extensivity of Solar Wind to be one more case of the universality of Tsallis theory in an extended series of various and discrete physical systems.

In this study we extend the phenomenological description and theoretical explanation of solar wind complexity by studying ion fluxes during solar wind shock events, using high time resolution measurements obtained by the Czech-Russian plasma spectrometer  BMSW on satellite Spectr-R. In section two, we present theoretical concepts presupposed for the data analysis, as well as the basic algorithm for the study of the nonextensive character of solar wind plasma medium during the shock events. In section 3, we present the instrumentation of the very high data resolution collection, as well as the result of data analysis. In section 4 we present the summary of the data analysis results and the theoretical interpretation in section 5.

## 2. Theoretical Preliminaries

### 2.1 Non-equilibrium Thermodynamics and non-extensive statistical mechanics

Solar Wind is a far from equilibrium thermodynamical system in which the development of hierarchical, self-organized and long-correlated dynamical states is possible. The nonlinearity in magnetohydrodynamic (MHD) description of solar wind plasma induces a general dynamical scenario of nonlinear partial differential equations. Namely, as the control parameter changes the dynamical profile of the solutions can bifurcate from simple self-organized states, such as limit cycles or torus, to strange attractors and multifractal spatiotemporal patterns. Also, the nonlinearity in solar wind dynamics can generate intermittent turbulence with the typical characteristics of the anomalous diffusion process and strange topologies of stochastic scalar wind velocity and magnetic fields are caused by the strange dynamics and strange kinetics.

Chang (1993, 1998, 2009) developed a powerful technique which utilizes scale invariance and the generalized renormalization group theory, used in critical non-equilibrium phenomena, for explaining self-organization and non-equilibrium phase transition processes in space plasmas. The results of this study testify this theoretical framework in the case of solar wind plasma, extended by the non-extensive statistical theory of Tsallis and the Tsallis $q$-entropy principle and the phase transition processes during solar wind shock events. This extended framework be used to describe the solar wind intermittent turbulence. In the next two sections we present the theoretical description of solar wind intermittent turbulence and the non-extensive statistical mechanics description of solar wind dynamics. From physical point of view, the non-equilibrium attracting complex states with long range spatial and temporal correlations correspond to non-equilibrium critical stationary states in which the entropy and free energy of the system become locally extremized.

### 2.1.1 Strange attractors and Self-Organization

When the dynamics is strongly nonlinear then, far from equilibrium, strong self-organization and intensive reduction of dimensionality of the state space can take place, caused by an attracting low dimensional set with parallel development of long range correlations in space and time. The attractor can be periodic (limit cycle, limit $m$-torus), simply chaotic (mono-fractal) or strongly chaotic with multiscale and multifractal profile as well as attractors with weak chaotic profile known as self organized critical (SOC) states. This spectrum of distinct dynamical profiles can be obtained as distinct critical points (critical states) of the nonlinear dynamics, after successive bifurcations as the control parameters change. The fixed points can be estimated by using a far from equilibrium renormalization procedure as it was proposed by Chang [Chang, 1992].

From this point of view phase transition processes can be developed between different critical states, when the order parameters of the system are changing. The far from equilibrium development of (weak or strong) chaotic critical states include long range correlations and multiscale internal self organization. The far from equilibrium self organized states, cause the equilibrium BG statistics and BG entropy, to transform by the Tsallis extension of $q-$statistics and Tsallis $q$-entropy. The extension of renormalization group theory and critical dynamics, under the $q-$extension of partition function, free energy and path integral approach was also proposed (Tsallis, 2009). The multifractal structure of the chaotic attractors can be described by the generalized Rényi fractal dimensions:

$$ D_{\bar{q}} = \frac{1}{\bar{q}-1} \lim_{\lambda \to 0} \frac{\log \sum_{i=1}^{N\lambda} p_i^{\bar{q}}}{\log \lambda}, \tag{1} $$

where $p_i \sim \lambda^{\alpha(i)}$ is the local probability at the location ($i$) of the phase space, $\lambda$ is the local size of phase space and $a(i)$ is the local fractal dimension of the dynamics. The Rényi $\bar{q}$ numbers (different from the $q-$index of Tsallis statistics) take values in the entire region $(-\infty, +\infty)$ of real numbers. The spectrum of distinct local pointwise dimensions $\alpha(i)$ is given by the estimation of the function $f(\alpha)$ defined by the scaling of the density $n(a, \lambda) \sim \lambda^{-f(a)}$, where $n(a, \lambda)da$ is the number of local regions that have a scaling index between $a$ and $a+da$. This reveals $f(a)$ as the fractal dimension of points with scaling index $a$. The fractal dimension $f(a)$ which varies with $a$ shows the multifractal character of the phase space dynamics which includes interwoven sets of singularity of strength $a$, by their own fractal measure $f(a)$ of dimension [Halsey,1986; Theiler,1990]. The multifractal spectrum $D_{\bar{q}}$ of the Renyi dimensions can be related to the spectrum $f(a)$ of local singularities by using the following relations:

$$ \sum p_i^{\bar{q}} = \int d\alpha' \, p(\alpha')\lambda^{-f(\alpha')}d\alpha' \tag{2} $$

$$ \tau(\bar{q}) \equiv (\bar{q}-1)D\bar{q} \stackrel{\min}{=} \bar{q}\alpha - f(\alpha) \tag{3} $$

$$ a(\bar{q}) = \frac{d[\tau(\bar{q})]}{d\bar{q}} \tag{4} $$

$$ f(\alpha) = \bar{q}\alpha - \tau(\bar{q}) \tag{5} $$

The physical meaning of these quantities included in relations (2-5) can be obtained if we identify the multifractal attractor as a thermodynamical object, where its temperature ($T$), free energy ($F$), entropy ($S$) and internal energy ($U$) are related to the properties of the multifractal attractor as follows:

$$\left.\begin{array}{ll} \overline{q} \Rightarrow \dfrac{1}{T}, & \tau(\overline{q}) = (\overline{q}-1)D_q \Rightarrow F \\[2mm] \alpha \Rightarrow U, & f(\alpha) \Rightarrow S \end{array}\right\} \qquad (6)$$

This correspondence presents the relations (4-6) as a thermodynamical Legendre transform [Paladin,1987]. When $\overline{q}$ increases to infinite ($+\infty$), which means, that we freeze the system ($T_{(q=+\infty)} \to 0$), then the trajectories (fluid lines) are closing on the attractor set, causing large probability values at regions of low fractal dimension, where $\alpha = \alpha_{min}$ and $D_{\overline{q}} = D_{+\infty}$. Oppositely, when $\overline{q}$ decreases to infinite ($-\infty$), that is we warm up the system ($T_{(q=-\infty)} \to 0$) then the trajectories are spread out at regions of high fractal dimension ($\alpha \Rightarrow \alpha_{max}$). Also for $\overline{q}' > \overline{q}$ we have $D_{\overline{q}'} < D_{\overline{q}}$ and $D_{\overline{q}} \Rightarrow D_{+\infty}(D_{-\infty})$ for $\alpha \Rightarrow \alpha_{min}(\alpha_{max})$ correspondingly. It is also known the Renyi's generalization of entropy according to the relation: $S_q^R = \dfrac{1}{q-1} \log \sum_i P_i^q$. However, the above description presents only a weak or limited analogy between multifractal and thermodynamical objects. The real thermodynamical character of the multifractal objects and multiscale dynamics was discovered after the definition by Tsallis (Tsallis, 2009) of the $q-$entropy related with the $q-$statistics as it is summarized in the next section. As Tsallis has shown Renyi's entropy as well as other generalizations of entropy cannot be used as the base of the non-extensive generalization of thermodynamics.

### 2.1.2 The Highlights of Tsallis Theory.

As we showed in a previous study (Pavlos, 2013) in many physical systems we can verify the presence of Tsallis statistics. This discovery is the continuation of a more general ascertainment of Tsallis $q$-extensive statistics from the macroscopic to the microscopic level (Tsallis, 2009). In our understanding the Tsallis theory, is not only a simple generalization of thermodynamics for chaotic and complex systems, or a non-equilibrium generalization of B-G statistics, but also a powerful theoretical foundation for the unification of macroscopic and microscopic physical complexity. From this point of view, Tsallis statistical theory is the other side of the modern fractal generalization of dynamics, while its essence is the efficiency of self-organization and development of long range correlations of coherent structures in complex systems.

From a general philosophical aspect, the Tsallis $q$-extension of statistics can be identified with the activity of an ordering principle in physical reality, which cannot be exhausted with the local interactions in the physical systems, as we mentioned in previous sections.

### 2.1.3 The non-extensive entropy ($S_q$).

Usually any extension of physical theory is related to some special form of mathematics. Tsallis non-extensive statistical theory is connected with the $q$-extension of exponential – logarithmic functions as well as the $q$-extension of Fourier Transform (FT). The $q$-extension of mathematics underlying the $q$-extension of statistics, are included in the solution of the non-linear equation:

$$\frac{dy}{dx} = y^q, (y(0) = 1, q \in R) \tag{7}$$

According to (7) its solution is the $q$-exponential function $e_q^x$:

$$e_q^x \equiv \left[1 + (1-q)x\right]^{1/(1-q)} \tag{8}$$

The $q$-extension of logarithmic function is the reverse of $e_q^x$ according to:

$$\ln_q x \equiv \frac{x^{1-q} - 1}{1 - q} \tag{9}$$

The $q$-logarithm satisfies the property:

$$\ln_q(x_A x_B) = \ln_q x_A + \ln_q x_B + (1-q)(\ln_q x_A)(\ln_q x_B) \tag{10}$$

In relation of the pseudo-additive property of the $q$-logarithm, a generalization of the product and sum to $q$-product and $q$-sum was introduced (7):

$$x \otimes_q y \equiv e_q^{\ln_q x + \ln_q y} \tag{11}$$

$$x \otimes_q y \equiv x + y + (1-q)xy \tag{12}$$

Moreover in the context of the $q$-generalization of the central limit theorem the $q$-extension of Fourier transform was introduced (1):

$$F_q[p](\xi) \equiv \int_{-\infty}^{+\infty} dx \, e_q^{ix\xi[p(x)]^{q-1}} p(x), \quad (q \geq 1) \tag{13}$$

It was for first time that Tsallis (Tsallis, 2009), inspired by multifractal analysis, conceived that the Boltzmann – Gibbs entropy

$$S_{BG} = -k \sum p_i \ln p_i = k < \ln(1 / p_i) > \tag{14}$$

is inefficient to describe the complex phenomenology of non-linear dynamical systems. The Boltzmann – Gibbs statistical theory presupposes ergodicity of the underlying dynamics in the system phase space. The complexity of dynamics, which is far different from the simple ergodic complexity, can be described by Tsallis non-extensive statistics, based on the extended concept of $q$ – entropy:

$$S_q = k \left( 1 - \sum_{i=1}^{N} p_i^q \right) / \left( q - 1 \right) = k < \ln_q (1 / p_i) > \qquad (15)$$

while for continuous state space, we have

$$S_q = k \left[ 1 - \int \left[ p(x) \right]^q dx \right] / \left( q - 1 \right) \qquad (16)$$

For a system of particles and fields with short range correlations inside their immediate neighbourhood, the Tsallis $q$ − entropy $S_q$ asymptotically leads to Boltzmann − Gibbs entropy ($S_{BG}$) corresponding to the value of $q = 1$. For probabilistically dependent or correlated systems $A, B$ it can be proven that:

$$\begin{aligned} S_q(A+B) &= S_q(A) + S_q(B / A) + (1-q) S_q(A) S_q(B / A) \\ &= S_q(B) + S_q(A / B) + (1-q) S_q(B) S_q(A / B) \end{aligned} \qquad (17)$$

where $S_q(A) \equiv S_q \left( \left\{ p_i^A \right\} \right), S_q(B) \equiv Sq \left( \left\{ p_i^B \right\} \right), S_q(B / A)$ and $S_q(A / B)$ are the conditional entropies of systems $A, B$. When the systems are probabilistically independent, then relation (17) is transformed to:

$$S_q(A+B) = S_q(A) + S_q(B) + (1-q) S_q(A) S_q(B) \qquad (18)$$

The first part of $S_q(A+B)$ is additive ($S_q(A) + S_q(B)$) while the second part is multiplicative including long − range correlations supporting the macroscopic ordering phenomena. Milovanov and Zelenyi (2000) showed that the Tsallis definition of entropy coincides with the so-called "kappa" distribution which appears in space plasmas and other physical realizations. Also, they indicate that the application of Tsallis entropy formalism corresponds to physical systems whose the statistical weights are relatively small, while for large statistical weights the standard statistical mechanism of Boltzmann − Gibbs is better. This result means that when the dynamics of the system is attracted in a confined subset of the phase space, then long − range correlations can be developed. Also according to Tsallis if the correlations are either strictly or asymptotically inexistent the Boltzmann − Gibbs entropy is extensive whereas $S_q$ for $q \neq 1$ is non extensive.

### 2.1.4 . The q- extension of statistics and Thermodynamics

According to the Tsallis $q$-extension of the entropy principle, any stationary random variable can be described as the stationary solution of generalized fraction of diffusion equation. At metastable stationary solutions of a stochastic process, the maximum entropy principle of Boltzmann − Gibbs statistical theory can faithfully be described the maximum (extreme) of the Tsallis $q$-entropy function. The extremization of Tsallis $q$-entropy corresponds to the $q$-generalized form of the normal distribution function:

$$p_q(x) = A_q \sqrt{\beta} e_q^{-\beta(x<x>_q)^2} \tag{19}$$

where $A_q = \sqrt{(q-1)/\pi}\,\Gamma(1/(q-1))/\Gamma((3-q)/[2/(q-1)])$ for $q > 1$,

and $A_q = \sqrt{(1-q)/\pi}\,\Gamma((5-3q)/[2(1-q)])/\Gamma((2-q)/(1-q))$ for $q < 1$, $\Gamma(z)$ being the Riemann function.

The $q$-extension of statistics includes the $q$-extension of central limit theorem which can describe also faithfully the non-equilibrium long range correlations in a complex system. The normal central limit theorem concerns Gaussian random variables $(x_i)$ for which their sum $Z = \sum_{i=1}^{N} x_i$ gradually tends to become a Gaussian process as $N \to \infty$, while its fluctuations tend to zero in contrast to the possibility of non-equilibrium long range correlations. By using q-extension to Fourier transform (see section 5 in this study), it can proved that $q$-independence means statistical independence for $q = 1$ (normal central limit theorem), but for $q \neq 1$ it means strong correlation ($q$-extended central limit theorem). In this case $(q \neq 1)$ the number of $W_{A_1+A_2+...+A_N}$ of allowed states, in a composed system by the $(A_1, A_2, ..., A_N)$ sub-systems, is expected to be smaller than $W_{A_1+A_2+...+A_N} = \Pi_{i=1}^{N} W_{A_i}$ where $W_{A_1}, W_{A_2}, ..., W_{A_N}$ are the possible states of the subsystems.

In this way, Tsallis $q-$extension of statistical physics opened the road for the $q-$extension of thermodynamics and general critical dynamical theory as the non-linear system lives far from thermodynamical equilibrium. The generalization of Boltzmann-Gibbs nonequilibrium statistics to Tsallis nonequilibrium $q$-statistics can be obtained by following Binney [Binney,1992]. In the next we present $q$-extended relations, which can describe the non-equilibrium fluctuations and $n-$point correlation function ($G$) can be obtained by using the Tsallis partition function $Z_q$ of the system as follows:

$$G_q^n(i_1, i_2, ..., i_n) \equiv \left\langle s_{i_1}, s_{i_2}, ..., s_{i_n} \right\rangle_q = \frac{1}{z} \frac{\partial^n Z_q}{\partial j_{i_1} \cdot \partial j_{i_2} ... \partial j_{i_n}} \tag{20}$$

where $\{s_i\}$ are the dynamical variables and $\{j_i\}$ their sources included in the effective $-$ Lagrangian of the system. As correlation (Green) equations (20) describe discrete variables, the $n-$point correlations for continuous distribution of variables (random fields) are given by the functional derivatives of the functional partition as follows:

$$G_q^n(\vec{x}_1, \vec{x}_2, ..., \vec{x}_n) \equiv \left\langle \varphi(\vec{x}_1)\varphi(\vec{x}_2)...\varphi(\vec{x}_n) \right\rangle_q = \frac{1}{Z} \frac{\delta}{\delta J(\vec{x}_1)} ... \frac{\delta}{\delta J(\vec{x}_n)} Z_q(J) \tag{21}$$

where $\varphi(\vec{x})$ are random fields of the system variables and $j(\vec{x})$ their field sources. The connected $n-$point correlation functions $G_i^n$ are given by:

$$G_q^n(\vec{x}_1, \vec{x}_2, ..., \vec{x}_n) \equiv \frac{\delta}{\delta J(\vec{x}_1)}...\frac{\delta}{\delta J(\vec{x}_n)} \log Z_q(J) \qquad (22)$$

The connected $n-$point correlations correspond to correlations that are due to internal interactions defined as:

$$G_q^n(\vec{x}_1, \vec{x}_2, ..., \vec{x}_n) \equiv \left\langle \varphi(\vec{x}_1)...\varphi(\vec{x}_n) \right\rangle_q - \left\langle \varphi(x_1)...\varphi(x_n) \right\rangle_q \qquad (23)$$

The probability of the microscopic dynamical configurations is given by the general relation:

$$P(conf) = e^{-\beta S_{conf}} \qquad (24)$$

where $\beta = 1/k\mathrm{T}$ and $S_{conf}$ is the action of the system, while the partition function $Z$ of the system is given by the relation:

$$Z = \sum_{conf} e^{-\beta S_{conf}} \qquad (25)$$

The $q-$extension of the above statistical theory can be obtained by the $q-$partition function $Z_q$. The $q-$partition function is related with the meta-equilibrium distribution of the canonical ensemble which is given by the relation:

$$p_i = e_q^{-\beta q(E_i - V_q)/Z_q} \qquad (26)$$

with

$$Z_q = \sum_{conf} e_q^{-\beta q(E_i - V_q)} \qquad (27)$$

and

$$\beta_q = \beta / \sum_{conf} p_i^q \qquad (28)$$

where $\beta = 1/KT$ is the Lagrange parameter associated with the energy constraint:

$$\left\langle E \right\rangle_q \equiv \sum_{conf} p_i^q E_i / \sum_{conf} p_i^q = U_q \qquad (29)$$

The $q-$extension of thermodynamics is related to the estimation of $q-$free energy $(F_q)$, the $q-$expectation value of internal energy $(U_q)$, the $q-$specific heat $(C_q)$ by using the $q-$partition function:

$$F_q \equiv U_q - TS_q = -\frac{1}{\beta}\ln qZ_q \qquad (30)$$

$$U_q = \frac{\partial}{\partial \beta}\ln qZ_q, \frac{1}{T} = \frac{\partial S_q}{\partial U_q} \qquad (31)$$

$$C_q \equiv T\frac{\partial \delta_q}{\partial T} = \frac{\partial U_q}{\partial T} = -T\frac{\partial^2 F_q}{\partial T^2} \qquad (32)$$

## 2.2 Intermittent Turbulence

The fractal-multifractal structuring of phase space, caused by the non-linear dynamics includes islands, cantori and stickiness and is related to singular measures, singular (irregular) functions of space and time (fractal functions), as well as to scale invariance properties and multiscale interaction causing long range correlations and hierarchical structures (Schlesinger et al. 1987, 1993; Schlesinger 1988; Arneodo et al. 1995; Zalsavsky, 2002). The $q$-extended statistical mechanics and the Tsallis $q$-distributions correspond to general power law probability functions with local singularities ($\alpha$) related to the singularity spectrum functions $f(\alpha)$ and the generalized fractal dimension spectrum functions $D_q$ (Theiler, 1989; Arneodo et al., 1995). The smaller singularity strength $\alpha(x)$ at the point ($x$) causes stronger singularities around ($x$) with intensive multiscale interactions and self-organization process. The fractal-multifractal hierarchical structuring of the phase space is mirrored to the fractal-multifractal character of the stochastic physical fields (velocities, electro-magnetic fields etc.), in the physical space and time. The energy dissipation field ($\varepsilon_\ell$) and the velocity field ($U_\ell$) satisfy scaling laws such as:

$$\varepsilon_\ell(x) \sim \ell^{\alpha-1}, \, u_\ell \sim \ell^h, \ell \to 0 \qquad (33)$$

while the corresponding probability distribution functions follow also scaling laws:

$$P_\ell(\varepsilon_\ell) \sim \ell^{d-f(\alpha)}, \, P_\ell(U_\ell) \sim \ell^{d-D(h)}, \, \ell \to 0 \qquad (34)$$

Where $d$ is the dimension of the physical space, $f(\alpha)$ and $D(h)$ are the fractal dimensions of the fractal sets, corresponding to the singularities ($\alpha$, $h$) of the energy and velocity fields. The dissipated energy ($\varepsilon_\ell$) in the multifractal system of the turbulence field is a "mass" measure the moments of which are related to the partition function $Z(q, \varepsilon_\ell)$ and follow also a power law at small $\ell$:

$$Z(q, \varepsilon_\ell) \equiv \sum \left\langle \varepsilon_\ell^{\bar{q}} \right\rangle \sim \ell^{\tau(\bar{q})} \qquad (35)$$

The exponent $\tau(\overline{q})$ is known as "mass exponent" and is associated with the generalized fractal dimensions $D_{\overline{q}}$ of the system from the relation

$$\tau_d(\overline{q}) = (\overline{q}-1)D_{\overline{q}} + (d-1)\overline{q} \qquad (36)$$

The $\tau(\overline{q})$ spectrum and in turn the $D_{\overline{q}}$ are obtained by the Legendre transformation of the singularity spectrum $f(\alpha)$ as follows (Frisch, 1996):

$$\tau(\overline{q}) = \min_{\alpha}(\overline{q}\alpha - f(a))$$
$$\overline{q} = \frac{df(\alpha)}{d\alpha} \qquad (37)$$

The relations (33-34) describe the multifractal and multiscale turbulent process in the physical state. The relations (35-37) describe the multifractal and multiscale process on the attracting set of the phase space. By using experimental timeseries we can construct the function $D_{\overline{q}}$ of the generalized Rényi $d-$dimensional space dimensions, while the relation (37) allows the calculation of the fractal exponent ($a$) and the corresponding multifractal spectrum $f_d(a)$. For homogeneous fractals of the turbulent dynamics the generalized dimension spectrum $D_{\overline{q}}$ is constant and equal to the fractal dimension of the support (Halsey, 1986; Paladin, 1987; Ferri, 2010). Kolmogorov (1941) stated that $D_{\overline{q}}$ does not depend on $\overline{q}$ as the dimension of the fractal support is $D_q = 3$. In this case the multifractal spectrum consists of the single point ($a = 1$ and $f(1) = 3$). The singularities of degree ($a$) of the dissipated fields, fill the physical space of dimension $d$ with a fractal dimension $F(a)$, while the probability $P(a)da$ to find a point of singularity ($a$) is specified by the probability density $P(a)da \sim \ln^{d-F(a)}$. The filling space fractal dimension $F(a)$ is related to the multifractal spectrum function $f_d(a) = F(a) - (d-1)$, while, according to the distribution function $\Pi_{dis}(\varepsilon_n)$ of the energy transfer rate associated with the singularity $a$, corresponds also to the singularity probability as $\Pi_{dis}(\varepsilon_n)d\varepsilon_n = P(a)da$ (Arimitsu and Arimitsu, 2000).

Moreover the partition function $\sum_i P_i^{\overline{q}}$ of the Rényi fractal dimensions estimated by the experimental time series includes information for the local and global dissipation process of the turbulent dynamics as well as for the local and global dynamics of the attractor set, as it is transformed to the partition function $\sum_i P_i^q = Z_q$ of the Tsallis $q$-statistic theory.

In the following, we follow Arimitsu and Arimitsu (2000) for the theoretical estimation of significant quantitative relations which can also be estimated experimentally. The probability singularity distribution $P(a)$ can be estimated as extremizing the Tsallis entropy functional $S_q$. According to Arimitsu and Arimitsu

(2000) the extremizing probability density function $P(a)$ is given as a $q$–exponential function:

$$P(a) = Z_q^{-1}[1-(1-q)\frac{(a-a_0)^2}{2X/\ln 2}]^{\frac{1}{1-q}} \qquad (38)$$

where the partition function $Z_q$ is given by the relation:

$$Z_q = \sqrt{2X/[(1-q)\ln 2]} \ \ B(1/2, 2/1-q) \qquad (39)$$

and $B(a,b)$ is the Beta function. The partition function $Z_q$ as well as the quantities $X$ and $q$ can be estimated using the following equations:

$$\left. \begin{array}{c} \sqrt{2X} = \left[\sqrt{a_0^2+(1-q)^2}-(1-q)\right]/\sqrt{b} \\ b = (1-2^{-(1-q)})/[(1-q)\ln_2] \end{array} \right\} \qquad (40)$$

Consequently, the exponent's spectrum $f(a)$ can be found using the relation $P(a) \approx \ln^{d-F(a)}$ as follows:

$$f(a) = D_0 + \log_2[1-(1-q)\frac{(a-a_o)^2}{2X/\ln 2}]/(1-q)^{-1} \qquad (41)$$

where $a_0$ corresponds to the $q$–expectation (mean) value of $a$ through the relation:

$$<(a-a_0)^2>_q = (\int da P(a)^q (a-a_0)^q)/\int da P(a)^q \qquad (42)$$

while the $q$–expectation value $a_0$ corresponds to the maximum of the function $f(a)$ as $df(a)/da \mid a_0 = 0$. For the Gaussian dynamics ($q \to 1$) we have mono-fractal spectrum $f(a_0) = D_0$. The mass exponent $\tau(\overline{q})$ can be also estimated by using the inverse Legendre transformation: $\tau(\overline{q}) = a\overline{q} - f(a)$ as follows:

$$\tau(\overline{q}) = \overline{q}a_0 - 1 - \frac{2X\overline{q}^2}{1+\sqrt{C_{\overline{q}}}} - \frac{1}{1-q}[1-\log_2(1+\sqrt{C_{\overline{q}}})] \qquad (43)$$

Where $C_{\overline{q}} = 1 + 2\overline{q}^2(1-q)X\ln 2$.

The relation between $a$ and $q$ can be found by solving the Legendre transformation equation $\overline{q} = df(a)/da$. Also if we use the equations (41-43) we can obtain the relation:

$$a_{\overline{q}} - a_0 = (1-\sqrt{C_{\overline{q}}})/[\overline{q}(1-q)\ln 2] \qquad (44)$$

The $q-$index is related to the scaling transformations of the multifractal nature of turbulence according to the relation $q = 1 - a$. Arimitsu and Arimitsu [Arimitsu,2000] estimated the $q-$index by analyzing the fully developed turbulence state in terms of Tsallis statistics as follows:

$$\frac{1}{1-q} = \frac{1}{a_-} - \frac{1}{a_+}$$

(45)

where $a_\pm$ satisfy the equation $f(a_\pm) = 0$ of the multifractal exponents spectrum $f(a)$. This relation can be used for the estimation of $q_{sen}-$index included in the Tsallis $q-$triplet (see next section).

In order to determine the three parameters $X, q$ and $a_o$ we need three independent equations which are given:

$$\tau(1) = 0, \mu = 1 + \tau(2), \frac{1}{1-q} = (\alpha_-)^{-1} - (\alpha_+)^{-1}$$

(46)

where $\mu$ is the intermittency exponent refer to the scaling exponent of the autocorrelation function of energy dissipation:

$$\left\langle \varepsilon_l^2 \right\rangle \sim l^{-\mu}$$

(47)

Also $\mu$ is related to the curve $D_{\overline{q}}$ in the multifractal formalism according to Milovanov and Sreenivasan (1990):

$$\mu = -d^2 \left[ (\overline{q} - 1) D_{\overline{q}} \right] / d\overline{q}^2 \Big|_{\overline{q} = 0}$$

(48)

According to Meneveau and Sreenivasan (1990) the intermittency exponent $\mu$ for lognormal distributions is related to the multifractal spectrum as follows:

$$f(\alpha) = [d - (\alpha - \alpha_0)^2] / (2\mu)$$
$$\mu = 2(\alpha_0 - \alpha)$$

(49)

## 2.5 The $q$- triplet of Tsallis

The non-extensive statistical theory is based mathematically on the nonlinear equation:

$$\frac{dy}{dx} = y^q, \ ( \, y(0) = 1, q \in \Re \, )$$

(50)

with solution the $q$–exponential function such as: $e_q^x = [1 + (1-q)x]^{\frac{1}{1-q}}$. The solution of this equation can be realized in three distinct ways included in the $q$–triplet of Tsallis: ($q_{sen}, q_{stat}, q_{rel}$). These quantities characterize three physical processes which are summarized here, while the $q$–triplet values characterize the attractor set of the dynamics in the phase space of the dynamics and they can change when the dynamics of the system is attracted to another attractor set of the phase space. The equation (50) for $q = 1$ corresponds to the case of equilibrium Gaussian Boltzmann-Gibbs (BG) world (Tsallis, 2009). In this case of equilibrium BG world the $q$–triplet of Tsallis is simplified to ($q_{sen} = 1, q_{stat} = 1, q_{rel} = 1$). The deeper theoretical foundation of Tsallis q-triplet is presented in the section (5) of this study.

### 2.5.1. The $q_{stat}$ index and the non-extensive physical states

According to Tsallis (Tsallis,2009), the long range correlated metaequilibrium non-extensive physical process can be described by the nonlinear differential equation:

$$\frac{d(p_i Z_{stat})}{dE_i} = -\beta q_{stat}(p_i Z_{stat})^{q_{stat}} \tag{51}$$

The solution of this equation corresponds to the probability distribution:

$$p_i = e_{q_{stat}}^{-\beta_{stat} E_i} / Z_{q_{stat}} \tag{52}$$

where $\beta_{q_{stat}} = \dfrac{1}{K T_{stat}}$, $Z_{stat} = \sum_j e_{q_{stat}}^{-\beta q_{stat} E_j}$.

Then the probability distribution function is given by the relations:

$$p_i \propto \left[1 - (1-q)\beta_{q_{stat}} E_i\right]^{1/1-q_{stat}} \tag{53}$$

for discrete energy states $\{E_i\}$ by the relation:

$$p(x) \propto \left[1 - (1-q)\beta_{q_{stat}} x^2\right]^{1/1-q_{stat}} \tag{54}$$

for continuous $x$ states of $\{X\}$, where the values of the magnitude $X$ correspond to the state points of the phase space.

The above distributions functions (53,54) correspond to the attracting stationary solution of the extended (anomalous) diffusion equation related to the nonlinear dynamics of the system. The stationary solutions $P(x)$ describe the probabilistic character of the dynamics on the attractor set of the phase space. The non-equilibrium dynamics can be evolved on distinct attractor sets depending upon the control parameters values, while the $q_{stat}$ exponent can change as the attractor set of the dynamics changes.

### 2.5.2. The $q_{sen}$ index and the entropy production process

The entropy production process is related to the general profile of the attractor set of the dynamics. The profile of the attractor can be described by its multifractality as well as by its sensitivity to initial conditions. The sensitivity to initial conditions can be described as follows:

$$\frac{d\xi}{d\tau} = \lambda_1 \xi + (\lambda_q - \lambda_1)\xi^q \tag{55}$$

Where $\xi$ describes the deviation of trajectories in the phase space by the relation: $\xi \equiv \lim_{\Delta(x)\to 0}\{\Delta x(t) \setminus \Delta x(0)\}$ and $\Delta x(t)$ is the distance of neighbouring trajectories (Tsallis, 2002). The solution of equation (55) is given by:

$$\xi = \left[ 1 - \frac{\lambda q_{sen}}{\lambda_1} + \frac{\lambda q_{sen}}{\lambda_1} e^{(1-q_{sen})\lambda_1 t} \right]^{\frac{1}{1-q}} \tag{56}$$

The $q_{sen}$ exponent can be also related to the multifractal profile of the attractor set by the relation:

$$\frac{1}{q_{sen}} = \frac{1}{a_{\min}} - \frac{1}{a_{\max}} \tag{57}$$

where $a_{\min}(a_{\max})$ corresponds to the zero points of the multifractal exponent spectrum $f(a)$ (Arimitsu & Arimitsu, 2001; Tsallis 2002, 2009). That is $f(a_{\min}) = f(a_{\max}) = 0$.

The deviations of neighbouring trajectories as well as the multifractal character of the dynamical attractor set in the system phase space are related to the chaotic phenomenon of entropy production according to Kolmogorov – Sinai entropy production theory and the Pesin theorem (Tsallis, 2009). The $q$ – entropy production is summarized in the equation:

$$K_q \equiv \lim_{t\to\infty} \lim_{W\to\infty} \lim_{N\to\infty} \frac{<S_q>(t)}{t} \tag{58}$$

The entropy production ($dS_q / t$) is identified with $K_q$, as $W$ are the number of non-overlapping little windows in phase space and $N$ the state points in the windows according to the relation $\sum_{i=1}^{W} N_i = N$. The $S_q$ entropy is estimated by the probabilities $P_i(t) \equiv N_i(t) / N$. According to Tsallis the entropy production $K_q$ is finite only for $q = q_{sen}$ (Tsallis, 2009).

### 2.5.3. The $q_{rel}$ index and the relaxation process

The thermodynamical fluctuation – dissipation theory (Chame and Mello, 1994) is based on the Einstein original diffusion theory (Brownian motion theory). Diffusion process is the physical mechanism for extremization of entropy. If $\Delta S$ denote the deviation of entropy from its equilibrium value $S_0$, then the probability of the proposed fluctuation that may occur is given by:

$$P \sim \exp(\Delta S / k) \tag{59}$$

The Einstein – Smoluchowski theory of Brownian motion was extended to the general Fokker – Planck diffusion theory of non-equilibrium processes. The potential of Fokker – Planck equation may include many metaequilibrium stationary states near or far away from the basic thermodynamical equilibrium state. Macroscopically, the relaxation to the equilibrium stationary state of some dynamical observable $O(t)$ related to the evolution of the system in phase space can be described by the form of general equation as follows:

$$\frac{d\Omega}{dt} \simeq -\frac{1}{\tau}\Omega \tag{60}$$

where $\Omega(t) \equiv [O(t) - O(\infty)] / [O(0) - O(\infty)]$ is the relaxing relevant quantity of $O(t)$ and describes the relaxation of the macroscopic observable $O(t)$ relaxing towards its stationary state value. The non-extensive generalization of fluctuation – dissipation theory is related to the general correlated anomalous diffusion processes (Tsallis, 2009). Now, the equilibrium relaxation process is transformed to the metaequilibrium non-extensive relaxation process

$$\frac{d\Omega}{dt} = -\frac{1}{T_{q_{rel}}}\Omega^{q_{rel}} \tag{61}$$

the solution of this equation is given by:

$$\Omega(t) \simeq e_{q_{rel}}^{-t/\tau_{rel}} \tag{62}$$

The autocorrelation function $C(t)$ or the mutual information $I(t)$ can be used as candidate observables $\Omega(t)$ for the estimation of $q_{rel}$. However, in contrast to the linear profile of the correlation function, the mutual information includes the non linearity of the underlying dynamics and it is proposed as a more faithful index of the relaxation process and the estimation of the Tsallis exponent $q_{rel}$.

## 3. Data Analysis and Results

### 3.1 The methodology of time series analysis

In the following first we describe the mathematical framework concerning the algorithm used for the solar wind time series analysis, according to the theoretical description of the previous section and then we provide the results of the analysis.

### 3.1.1 Flatness Coefficient $F$

The intermittent nature of the solar wind plasma dynamics can be investigated through the Probability Density Functions (PDF) of a set of two-point differenced time series of an original time series $\delta B_{\tau}(t) = B(t + \tau) - B(t)$, which can be any physical quantity. The coefficient $F$ corresponding to the flatness values of the two-point difference for the observed time series is defined as:

$$F = \frac{< \delta B_{\tau}(t)^4 >}{< \delta B_{\tau}(t)^2 >^2} \tag{63}$$

The coefficient $F$ for a Gaussian process is equal to 3. Deviation from Gaussian distributions implies intermittency, while the parameter $\tau$ represents the spatial size of the plasma eddies, which contribute to the energy cascade process. According to general theory of turbulence, intermittency appears in the heavy tails of the distribution functions as the dynamics in the vortex is non-random, but deterministic (Voros et al, 2004).

The flatness coefficient reveals the non-Gaussian character of the statistics but it cannot give more information for the higher than 4 moments of the distribution. In contrast to the flatness coefficient the Tsallis $q$-statistics and the structure function scaling exponents spectrum include much stronger information about the non-Gaussian character of the turbulent state than the flatness coefficient. Moreover, Tsallis theory can be used for the quantitative prediction of the multifractal character of the turbulent state making possible the comparison of theoretical predictions and experimental estimations.

### 3.1.2 The $q$-triplet estimation

According to previous analysis concerning the $q$- triplet of Tsallis, we estimate the ($q_{sen}, q_{stat}, q_{rel}$) as follows:

a) The $q_{sen}$ index is given by the relation:

$$q_{sen} = 1 + \frac{a_{max} a_{min}}{a_{max} - a_{min}} \qquad (64)$$

The $a_{max}, a_{min}$ values correspond to the zeros of multifractal spectrum function $f(a)$, which is estimated by the Legendre transformation $f(a) = \overline{q}a - (\overline{q} - 1)D_{\overline{q}}$, where $D_{\overline{q}}$ describes the Rényi generalized dimension of the solar wind time series in accordance with the relation:

$$D_{\overline{q}} = \frac{1}{q-1} \cdot \lim \left( \log \sum p_i^q \Big/ \log r \right) \text{for } r \to 0 \qquad (65)$$

The $f(a)$ and $D(\overline{q})$ functions can be estimated experimentally by using the relations (64-65) and the underlying theory according to March and Tu (March,1997) and Consolini (Consolini,1996). The same functions can be estimated also theoretically by using the Tsallis $q$-entropy principle.

b) The $q_{stat}$ values are derived from the observed Probability Distribution Functions (PDF) according to the Tsallis $q$-exponential distribution:

$$PDF[\Delta z] \equiv A_q \left[ 1 + (q-1)\beta_q (\Delta z)^2 \right]^{\frac{1}{1-q}} \qquad (66)$$

where the coefficient $A_q$, $\beta_q$ denote the normalization constants and $q \equiv q_{stat}$ is the entropic or non-extensivity factor ($q_{stat} \leq 3$) related to the size of the tail in the distributions. Our statistical analysis is based on the algorithm as described in Ferri (Ferri,2010). We construct the $PDF[\Delta z]$ which is associated with the first difference $\Delta z = z_{n+1} - z_n$ of the experimental sunspot time series, while the $\Delta z$ range is subdivided into little ``cells'' (data binning process) of width $\delta z$, centered at $z_i$ so that one can assess the frequency of $\Delta z$-values that fall within each cell/bin. The selection of the cell-size $\delta z$ is a crucial step of the algorithmic process and its equivalent to solving the binning problem: a proper initialization of the bins/cells can speed up the statistical analysis of the data set and lead to a convergence of the algorithmic process towards the exact solution. The resultant histogram is being properly normalized and the estimated $q$-value corresponds to the best linear fitting to the graph $\ln_q(p(z_i)) \, vs \, z_i^2$, where $\ln_q(p(z_i))$ is the so-called $q$-logarithm: $\ln_q(x) = (x^{1-q} - 1)/(1-q)$. Our algorithm estimates for each $\delta_q = 0,01$ step the linear adjustment on the graph under scrutiny (in this case the $\ln_q(p(z_i)) \, vs \, z_i^2$ graph) by evaluating the associated correlation coefficient *(CC)*, while the best linear fit is considered to be the one maximizing the correlation coefficient. The obtained $q_{stat}$, corresponding to the best linear adjustment is then being used to compute the following equation:

$$G_q(\beta, Z) = \frac{\sqrt{\beta}}{C_q} e_q^{-\beta z^2} \qquad (67)$$

where $C_q = \sqrt{\pi} \cdot \Gamma(\frac{3-q}{2(q-1)}) / \sqrt{q-1} \cdot \Gamma(\frac{1}{q-1})$, $\quad 1 < q < 3$ for different $\beta$-values. Moreover, we select the $\beta$-value minimizing the $\sum_i [G_{q_{sstat}}(\beta, Z_i) - p(Z_i)]^2$, as proposed in Ferri (Ferri,2010).

c) Finally the $q_{rel}$ index is given by the relation: $q_{rel} = (s-1)/s$ where $s$ is the slope of the log-log plotting of mutual information $I(\tau)$ given in Fraser and Swinney (Fraser,1986) by the relation:

$$I(\tau) = -\sum_{x(i)} P(x(i)) \log_2 P(x(i)) - \sum_{x(i-\tau)} P(x(i-\tau)) \log_2 P(x(i-\tau)) +$$
$$\sum_{x(i)} \sum_{x(i-\tau)} P(x(i), x(i-\tau)) \log_2 P(x(i), x(i-\tau)) \qquad (68)$$

### 3.1.3 Estimation of singularity spectrum $f(a)$

The singularity spectrum $f(a)$ can be estimated by using the Legendre transformation:

$$f(a) = \bar{q}a - \tau(\bar{q}), \ a = \frac{d\tau(\bar{q})}{d\bar{q}} \qquad (69)$$

where $\tau(\bar{q})$ is the mass exponent related to the generalized dimension function $D(\bar{q})$ through the relation $\tau(\bar{q}) = (\bar{q}-1)D_{\bar{q}}$. We also follow Consolini et al. (Consolini, 1996) for estimating the scaling exponent function $\tau(\bar{q})$ corresponding to the experimental time series $Z(t_i)$ after the estimation of the corresponding function $\Gamma(\bar{q}, \Delta t)$ according to the relation

$$\Gamma(\bar{q}, \Delta t) = \sum_{A_i} P_i(\Delta t)^{\bar{q}} \approx (\Delta t)^{\tau(\bar{q})} \qquad (70)$$

where $P_i(\Delta t)$ is the probability coarse-grained weight for time segments $A_i$ of the time size $(\Delta t)$ of the experimental signal.

### 3.1.4 Theoretical Prediction of the singularity spectrum $f(a)$

According to Arimitsu and Arimitsu (Arimitsu, 2000; 2001) we can also estimate theoretically the singularity exponent spectrum $f(a)$ by using the relation $P(a) \approx \ln^{d-F(a)}$ as follows:

$$f(a) = D_0 + \log_2[1 - (1-q)\frac{(a-a_o)^2}{2X/\ln 2}]/(1-q)^{-1} \qquad (71)$$

where $a_0$ corresponds to the $q$-expectation (mean) value of $a$ through the relation:

$$< (a-a_0)^2 >_q = (\int da P(a)^q (a-a_0)^q)/\int da P(a)^q \qquad (72)$$

while the $q$-expectation value $a_0$ corresponds to the maximum of the function $f(a)$ as $df(a)/da \,|\, a_0 = 0$. For the Gaussian dynamics ($q \to 1$) we have mono-fractal spectrum $f(a_0) = D_0$.

For the estimation of the singular spectrum function $f(a)$ we need three parameters ($a_0, q, X$). According to Arimitsu and Arimitsu (Arimitsu, 2000) we can estimate these parameters by using the equations:

$$\tau(\overline{q} = 1) = 0 \,, \ \mu = 1 + \tau(\overline{q} = 2) \text{ and } \frac{1}{1-q} = \frac{1}{a_-} + \frac{1}{a_+} \qquad (73)$$

where $\mu$ is the intermittency exponent (Frisch, 1996).

In this study we estimate the triplet of parameters in two ways. In the first way we use the experimental values of ($a_0, a_+, a_-$), while in the second way we estimate the intermittency exponent ($\mu$) which is used for the theoretical estimation of $q_{sen.}$

### 3.1.5 $p$-model prediction of the singularity spectrum $f(a)$

According to Meneveau and Sreenivasan (Meneveau, 1987), the $p$-model is a one-dimensional model version of a cascade model of eddies, each breaking down into two new ones according to a generalized two-scale Cantor set with $l_1 = l_2 = \frac{1}{2}$. The $p$-model was introduced to account for the occurrence of intermittency in fully developed turbulence. The best nonlinear fit of the generalized dimension $D(\overline{q})$ function is represented by

$$D_{\overline{q}} = \log_2 \left[ p^{\overline{q}} + (1-p)^{\overline{q}} \right]/1 - \overline{q} \qquad (74)$$

Where $p$ is the probability of the fragmentation in the cascade process and $q$ is the moment order.

## 3.2 Description of the data – solar wind measurements by device BMSW with high time resolution

The BMSW instrument (Fast Monitor of the Solar Wind) was specially designed for measuring of the solar wind plasma parameters with very high time resolution (Safrankova et.al, 2013; Zastenker et. al., 2013). The value of ion flux and it's direction are measured with resolution equal to 31 ms. The full function of energy distribution (and accordingly all solar wind parameters as ion density, velocity and temperature) is measured with time resolution equal to 3 s. BMSW instrument (see the photo in Fig.1a) includes 6 sensors of Faraday cup type. Three of them are used for determination of ion flux vector – the value and two angles of its direction. Three other sensors are used for determination of solar wind parameters – bulk velocity, ion density and ion temperature (in isotropic approximation).

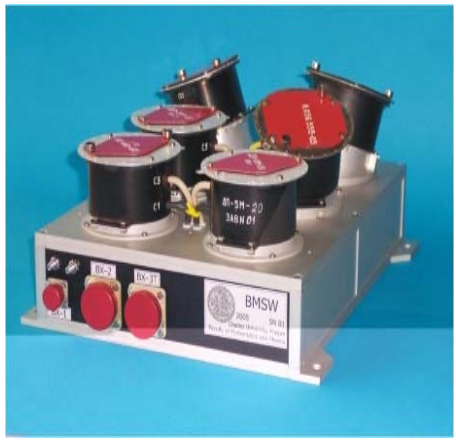 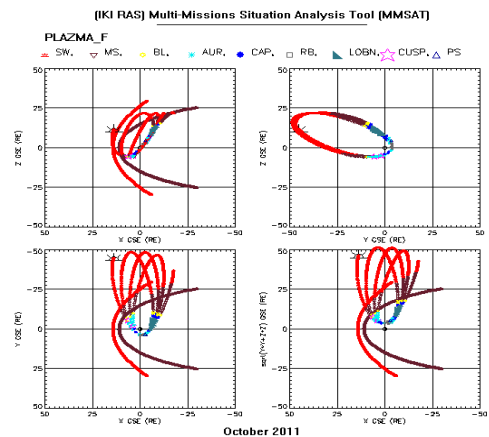

**Figure 1:** a) Photo of BMSW instrument, b) Typical SPECTR-R orbits evolution relative the Earth's magnetosphere in October 2011.

The BMSW instrument is installed onboard the high-apogee Spectr-R satellite. This spacecraft was launched on July 18, 2011, into orbit with apogee of ~ 350 000 km, perigee of ~ 50000 km and orbital period of 8.5 days. The BMSW instrument has operated almost permanently since August 6, 2011. The chosen orbit allows to carry out regular measurements of plasma parameters in both solar wind and Earth's magnetosheath in the March-November period of each year, during 7-8 days per each orbit. See examples of typical Spectr-R orbits evolution for one month relative the Earth's magnetosphere during the first year of flight in the Fig. 1b. The main axis of the BMSW instrument is permanently directed to the Sun within the limits of 5 –10 degrees' deviation. The instrument is equipped with its own Sun sensor which allows us to know the instrument's axes orientation relative to the Sun direction with accuracy of about 1 deg.

## 3.3 Experimental Time series

In this section we present results concerning three distinct shock events, labeled (A, B, C), in the solar wind plasma system observed in situ by the BMSW instrument. First, we present analytically the results of the experimental analysis for the event (C), while at the end of this section we summarize the results for all three shock events.

In Figure 2, the shock of 26 September 2011, is presented, for 5.5 hours time interval (09:40 – 15:00 UT) with 31 ms time resolution. The shock took place at ~ 12.35 UT. Panel «a» presents the ion flux value. At the moment of the shock front, ion flux grow from 5 up to $25*10^8$ $cm^{-2}$ $s^{-1}$. Below solar wind parameters for the same period are presented: ion density (panel b), ion velocity (panel c) and ion temperature (panel d). As we can see in this figure, during the calm period, namely from 9:50 till 12: 35 UT, the solar wind parameters attain low values with very small fluctuations. However, on shock font, there is a rapid increase of all solar wind parameters values. In particular, density shifts from 20 to 55 $cm^{-3}$, velocity from 360 to 460 km/s, temperature from 10 to 30 eV. Moreover, especially for ion flux and density there are also high valued rapid fluctuations. Thus, it's a typical example of interplanetary shock front.

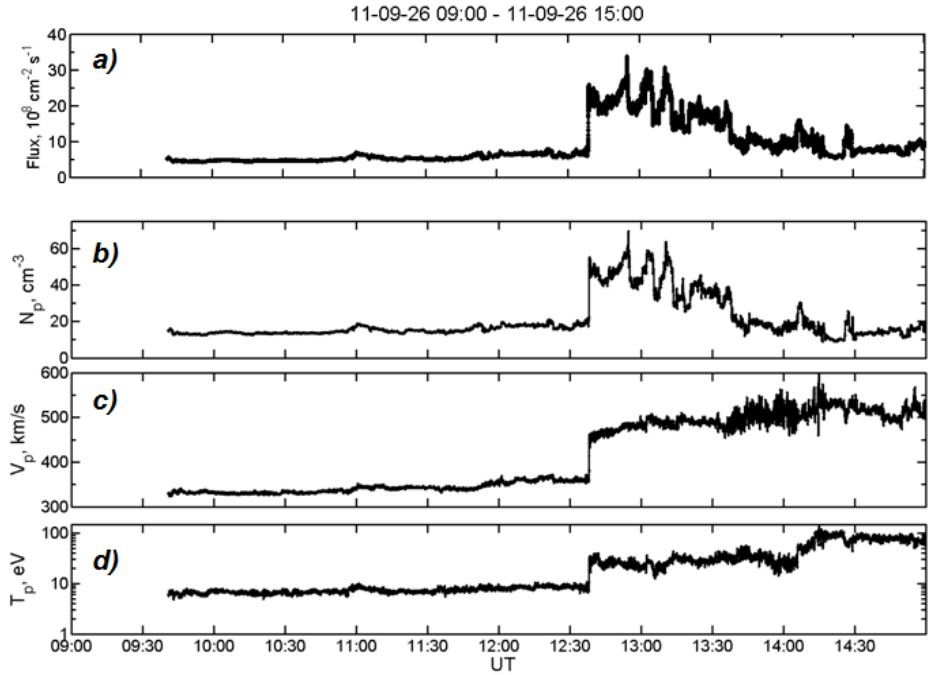

**Figure 2**: The Solar wind parameters Ion Flux (a), density (b), velocity (c), temperature (d) on shock front 26.09.2011

In order to study the evolution and the gradual phase transition of the dynamics of solar wind from calm to shock more thoroughly, we divided the original ion flux time series, consisting of 604.510 counts, into five segments shown in Fig. 3. The first three segments (x1calm, x2calm, x3calm) correspond to the calm period time series (previous to shock), while the other two to the main shock period (x1shock) and its relaxation (x2shock). Furthermore, as far as the calm time series segments is concerned, the x1calm segment correspond to the initial calm period whereas the following segments, namely x2calm and x3 calm, are periods of possible phase transition, where precursory main shock phenomena may take place.

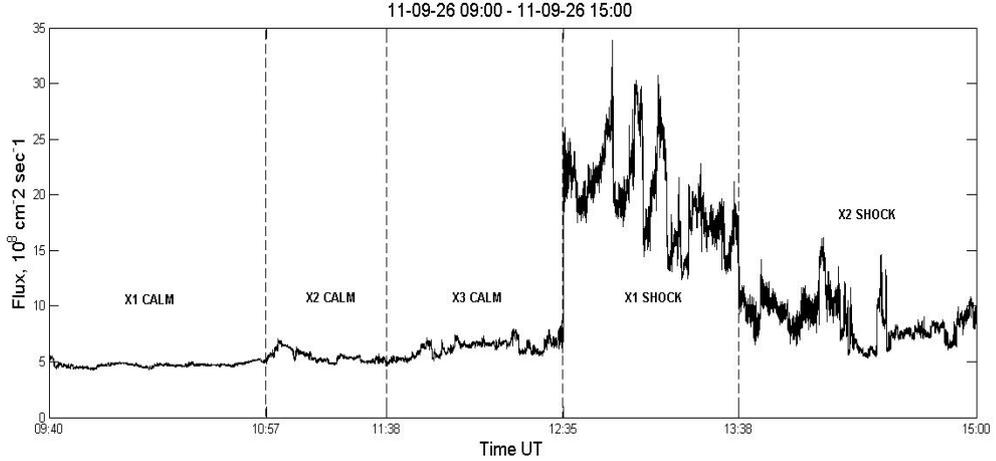

**Figure 3**: Time series of the ion flux divided in 5 segments, namely x1calm, x2calm, x3calm, x1shock, x2shock.

### 3.4 Flatness Coefficient

Figure 4 shows the flatness coefficient $F$, for $\tau = 1$, for the five time series segments. As it can be seen, the flatness coefficient $F$ for all the period, both for the calm and the shock, attains values much higher than 3 with a minimum of $F$=7.6 in the beginning of the calm period (x1calm) and a maximum $F$=19.6, in the end of the shock period (x2shock). This indicates the existence of non-Gaussian intermittent dynamics of the solar wind both in calm and shock period, while the significant increase of flatness coefficient $F$ reveals the stronger non-Gaussian intermittent turbulence character of the solar wind dynamics during shock period compared to the calm period. Furthermore, the flatness coefficient $F$ increases as the shock event approaches, attaining values approximately from 7.6 to 18. When the shock occurs (x1shock segment) the index $F$ decreases to value 15.5. The gradual increase of the coefficient $F$, implies the development of long range correlations till the shock occurs. When the shock occurs (x1shock), due to its bursting nature, rapid fluctuations evolve reducing the long range correlations but not to a significant degree. As the shock phenomenon relaxes the long range correlations re-establish (x2shock period). Finally, it must be noted, that the increase in flatness coefficient in the calm period, namely in x2calm and x3calm period, captures a change in the dynamics well before the shock period, that is indiscernible in the solar wind parameters shown in Fig. 1. This result could be explained considering the strong interaction that takes place between the upcoming fast solar wind and the, relatively steady, slow solar wind due to pressure effects. Furthermore, as we conjecture, this result can be connected with fracton dynamics which can cause oscillations of statistical parameters observed during shock events development.

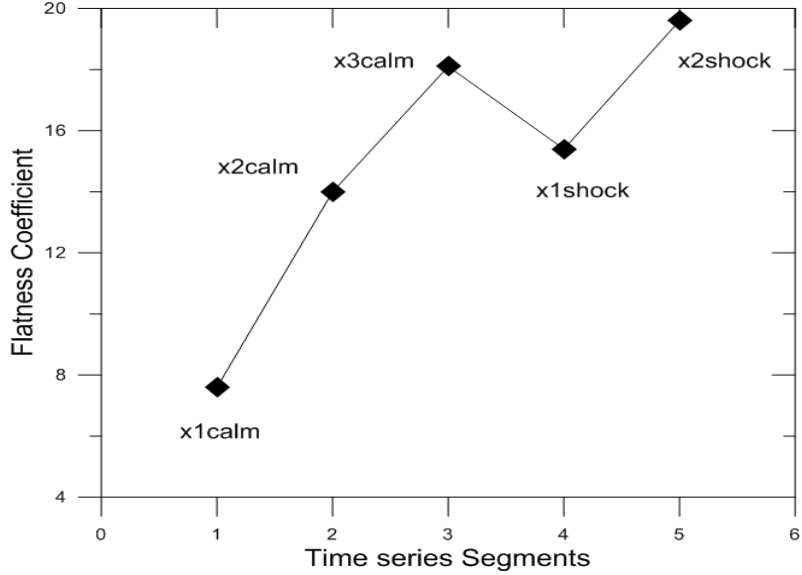

**Figure 4**: Flatness Coefficient $F$ estimated for the five ion flux time series of the Solar Wind.

### 3.5 Tsallis Entropy

Figure 5 shows the normalized mean values of Tsallis entropy $S_q$ (see Eq. 7) corresponding to each segment using the $q_{sens}$ index values (see below paragraph 3.6.1) and a window length $N = 100$. This length has been determined so as that the entropy essentially stops depending on the window size for sizes above this value (Gonzalez et al., 2011). The decrease of the entropy well before the onset of the shock main period, namely in x2calm segment, suggests a change in the organization level of the solar wind dynamics system connected with self organization processes and the development of a new emerging self organized state, with a lower complexity level, as evidenced by the loss of entropy. The entropy increases after the minimum, indicating a recovery of the complexity of the system. Moreover, the values of normalized Tsallis entropy reveal a small fluctuation around a lowest level, namely in x2calm, x3calm and x2shock segments. This result is in accordance with the results presented previously concerning flatness coefficient and can also be related to fracton dynamics.

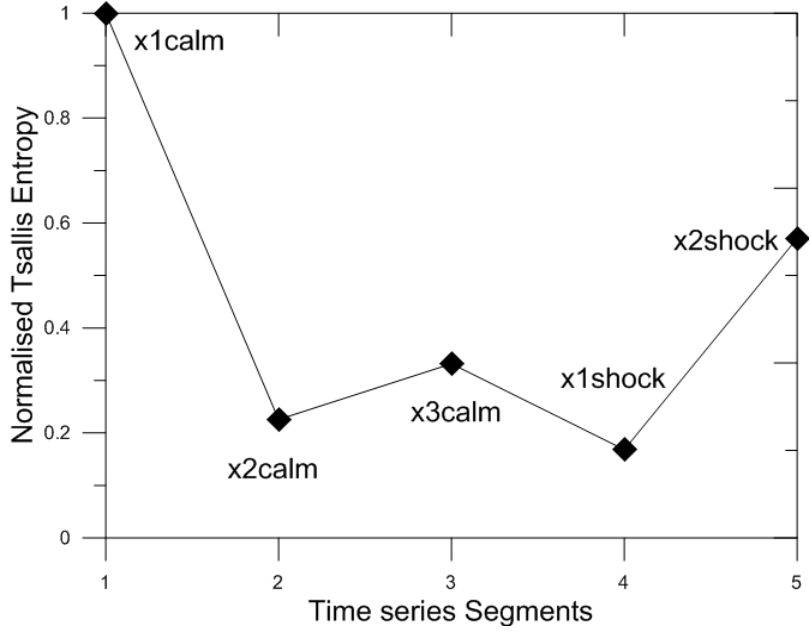

**Figure 5**: Normalized Tsallis entropy estimated for the five ion flux time series segments of the solar wind.

## 3.6 The non-extensive *q*-statistics of the Solar Wind Dynamics during Calm and Shock Period

Here, we present results concerning the estimation of $q_{sens}$, $q_{stat}$ and $q_{rel}$ indices for the solar wind ion flux time series during the quiet and shock period segments of Fig. 3.

### 3.6.1 Determination of $q_{sens}$

In the following we present the multifractal scaling spectra $f(a)$ and the corresponding dimension spectra $D_q$ estimated for the first calm and the first shock period, namely x1calm and x1shock segments of the solar wind ion flux time series. For the multifractal structure, which is a non-uniform fractal, the variation of $D_{\bar{q}}$ with $\bar{q}$ quantifies the non-uniformity. For large values of $\bar{q}$ (positive values) $D_{\bar{q}}$ is as associated with the infrequent points on the fractal and for small values of $\bar{q}$ (negative values) $D_{\bar{q}}$ is associated with dense points. For instance:

$$D_{\infty} = \lim_{\tau \to 0} \frac{\log(\max_i P_i)}{\log \tau}, \ D_{-\infty} = \lim_{\tau \to 0} \frac{\log(\min_i P_i)}{\log \tau} \qquad (75)$$

Namely, a region with dense point corresponds to high probability values $P_i$, small fractal dimensions ($D_{\bar{q}}$) and small scaling exponents ($\alpha$). In contrast regions with least dense points correspond to small probability values ($P_i$), large fractal dimensions ($D_{\bar{q}}$) and large scaling exponents ($\alpha$). The Legendre transformation between ($q,\tau$) and ($\alpha,f$) set of variables indicates the correspondence between $a_- = a_{min}$ , $D_{\infty}$ and $q = +\infty$, as well as the correspondence between $a_+ = a_{max}$, $D_{-\infty}$ and $q = -\infty$, as $D_{+\infty} = D(\bar{q} = +\infty)$ and $D_{-\infty} = D(\bar{q} = -\infty)$.

As we can notice in Figure 6 (a-d) the profile of the $f(a)$ and $D_q$ spectra changes drastically as the dynamics evolve from the calm to the shock active period. For the quantification of these changes, the $f(a)$ function was approximated with a fourth and sixth- degree polynomial correspondingly, so as to determine the $a_{min}$, $a_{max}$ values and their errors:

***Calm period***
$\Delta\alpha = \alpha_{max} - \alpha_{min} = 0.7525$, $\Delta D_q = D_{q=-\infty} - D_{q=+\infty} = 0.4311$

***Shock period***
$\Delta\alpha = \alpha_{max} - \alpha_{min} = 1.2049$, $\Delta D_q = D_{q=-\infty} - D_{q=+\infty} = 0.8709$

These results point to a strong multifractal behavior while the multifractal ranges $\Delta a, \Delta D_q$ increase from the calm to the shock period. The estimated $q_{sens}$ values by using the previous relation for the two periods were found to be: $q_{sens}(calm) = -0.2492 \pm 0.0001$, $q_{sens}(shock) = 0.30533 \pm 0.0125$. Since the $q_{sens}$ clearly increases as the dynamics shifts to the shock active period we can conclude the stronger instability of the system and the growth of $q$-entropy production, due to the increase of the multifractality of the phase space.

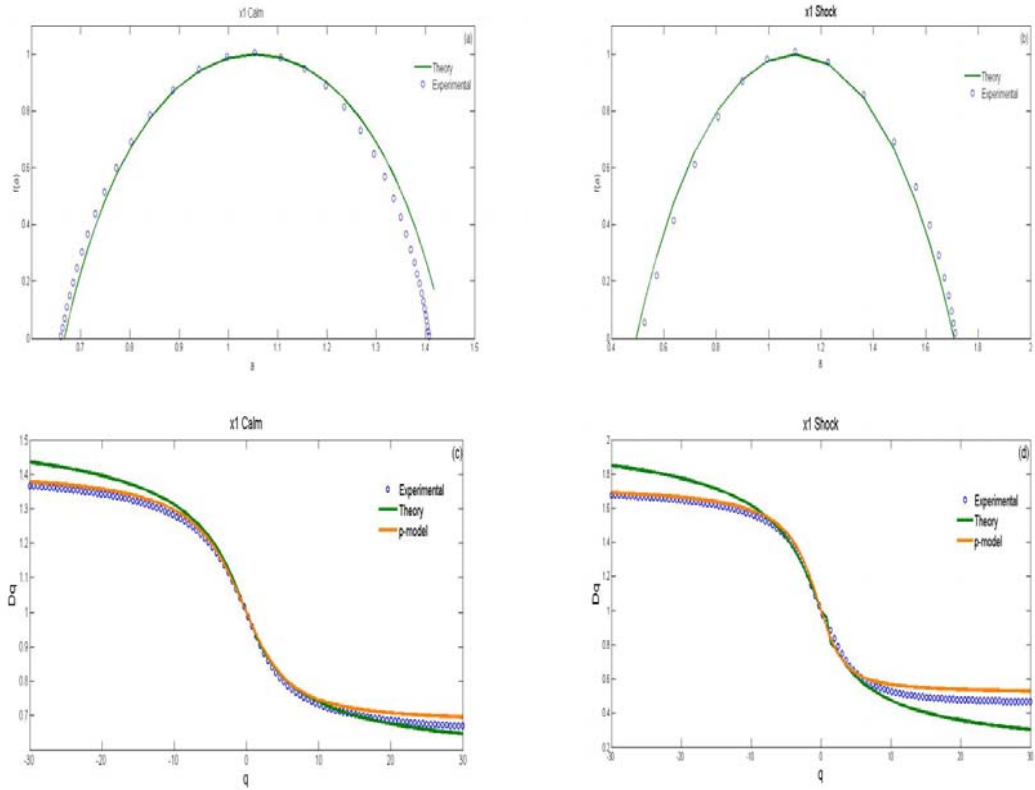

**Figure 6:** (a) Multifractal spectrum of solar wind ion flux calm period time series. The solid green line corresponds to theoretical estimation of the multifractal spectrum. The index was found to be $q_{sens}$= - $0.2422 \pm 0.0085$. (b) Same as (a) but for shock period time series. The estimation gave $q_{sens}$= $0.2731 \pm 0.0011$. (c) The generalized dimension $D(q)$ vs. $q$ of solar wind ion flux calm period time series. The solid green line corresponds to theoretical estimation of generalized dimension, the blue dots to experimental estimation and the orange line to $p$-model estimation (d) Same as (c) but for the solar wind ion flux shock period time series.

Furthermore, in Fig. 6, with green line we show the theoretical estimation of singularity spectrum $f(\alpha)$ and the generalized dimension spectrum $(D_{\tilde{q}})$, and with orange line the $p$-model prediction as far the generalized dimension is concerned. The methods are described analytically in previous sections 3.1.4 and 3.1.5 respectively. As we can see in Fig. 6a, corresponding to the calm period, the theoretical singularity spectrum fits very well the left part of the experimental singularity spectrum, while for the right part, there is a deviation. According to N. Arimitsu et al. [Arimitsu,2000] and T. Arimitsu et al. [Arimitsu,2001], the left-half of the experimentally estimated spectrum function $f(\alpha)$ corresponds to the tail-part of PDF for energy dissipation rates, and right-half of experimental $f(\alpha)$ is related to the center-part of PDF. However, the theoretical formula for the spectrum function is applicable only to the part of PDF constructed mainly from the intermittent singular element of turbulence. Since the tail-part of PDF is responsible for the intermittent singular behaviour of turbulent, the left-half of theoretical spectrum $f(\alpha)$ is the part which should be compared with the experimentally estimated spectral function. On the other hand, since the center part of PDF contains the element of thermal fluctuations originated from the dissipative term in Navier-Stokes equation in addition to the contribution from the singular element, the right-half of the theoretical spectrum $f(\alpha)$ is usually deviated from the experimentally estimated spectrum. Thus, in Fig. 6a the characteristics of $f(\alpha)$ given above are presenting quite well. The difference between the right-half theoretical spectrum $f(\alpha)$ and the right-half experimentally estimated, tells us in what degree the element of fluctuation contributes to the center-part of PDF compared with the intermittent singular element. On the other hand, Fig. 6b, corresponding to the shock period, shows that no such deviation in the right-half exists, namely the theoretical estimation fits quite well the experimental curve, indicating the efficiency of Tsallis theory in modelling the experimental data.

The theoretical estimated values of the generalized dimension $D(q)$ (green line) are presented at Figs. 6c,d, for the calm and shock period, correspondingly. As it is depicted in these figures the theoretical generalized dimension fits better the calm experimental generalized dimension compared to the shock experimental generalized dimension. In particular, there is a significant deviation in positive large values of $q$, for the shock period. According to Arimitsu et al. [Arimitsu,2000] for the function $D(q)$ the deviation between the theoretical and experimental $D(q)$ is due to the element of fluctuation contributing to the center-part of PDF. For larger positive $q$, the theoretical $D(q)$ may deviate from the experimental one. This is due to the cut-off of the data point of experimental PDF which is indispensable since the experimental PDF terminates itself at a certain tail point, i.e., causing the lack of the contribution from larger deviation. Finally, in the same figures the curve $p$-model (orange line) for modelling the experimental generalized dimension is shown. As it is shown, in both cases, for negative large and positive large values of $q$, the $p$-model is more efficient in fitting the experimental generalized dimension, than the theoretical estimation based on Tsallis theory. This result means that the underlying dynamics are connected with intermittency in fully developed turbulence.

In the following in Figure 7a we present the $\varDelta\alpha$ values corresponding to the five ion flux time series segments correspond to the densest $\left[a_{\min} = a\left(\overline{q} = +\infty\right)\right]$ regions of the attractor (similar results were obtained for the sparsest $\left[a_{\max} = a\left(\overline{q} = -\infty\right)\right]$ regions). As it is shown for the calm period there is a significant shift in multifractal range $\varDelta\alpha$ corresponding to x2calm segment. Continuously, there are fluctuations till there is a significant decrease of $\varDelta\alpha$ corresponding to x2shock segment.

Similar results were obtained from the valuation of $q_{sen}$ index values shown in Fig. 7b along with the bar errors. The $q_{\text{sen}}$ is related to entropy production of the system. As it is shown there is a remarkable increase in $q_{\text{sen}}$ value corresponding to x2calm period and x1shock, till there is a big decrease in x2shock segment. Thus, it seems that the system goes through a phase transition from a state of low entropy production (x1calm) to states of high entropy production (x2calm, x3calm, x1shock), while in the x2shock period recovers in a high degree the state of low entropy production. Similar results are obtained from the estimation of $p$-modeling values, shown in Fig. 7c.

The aforementioned results concerning the multifractal range $\varDelta\alpha$, the $q_{\text{sen}}$ index and the $p$-parameter are also in accordance with the results presented previously concerning the flatness coefficient and normalized Tsallis entropy and can also be related to fracton dynamics. These results also indicate the existence of a possible small shock event, possibly a precursor shock event, during the calm period, corresponding to the x2calm segment.

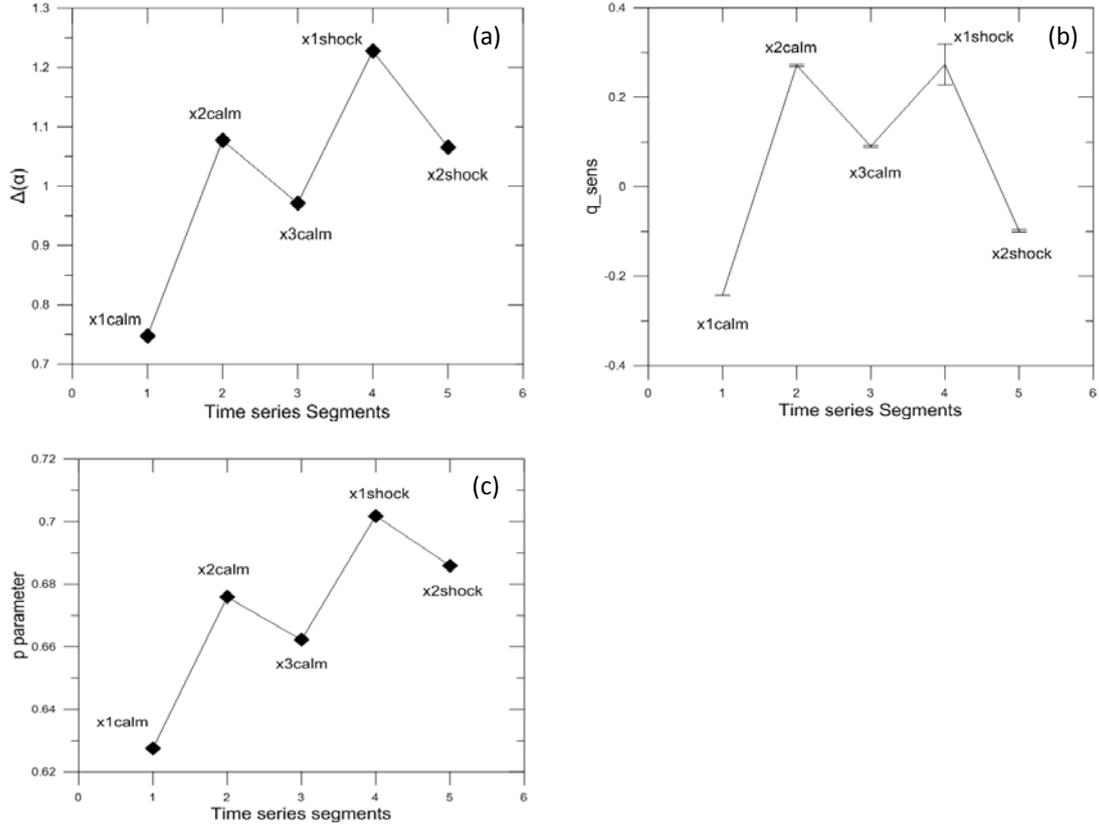

**Figure 7:** (a) Multifractal range, $\Delta\alpha$, for the five ion flux time series segments of the solar wind. (b) The $q_{sen}$ values for the five ion flux time series segments of the solar wind. c) The $p$ parameter values estimated for the five ion flux time series segments of the solar wind.

Furthermore, the application of Tsallis $q$-entropy principle can be used for the theoretical prediction of the values of intermittency coefficient $\mu$ and the $q_{sen}$ parameter in accordance with the theoretical description in section two. Table (1) presents the theoretically estimated values of parameters ($\alpha_0, q_{sen}, X$) by using the values of intermittency coefficient $\mu$. The values of $\mu$ were estimated in such a way so as to correspond to the experimentally estimated values of $\alpha_0$. Figures 8(a-c) present the variation of the $\mu$, $\alpha_0$ and $q_{sen}$ values estimated experimentally and theoretically for different periods during the shock events. In all cases, we can clearly observe a significant increase of the ($\mu, \alpha_0, q_{sen}$) corresponding to x2calm segment period, while there is a small fluctuation corresponding to x3calm and x1shock periods and finally a remarkable increase in x2shock period. Also, Fig. 8(d) presents the comparison of $q_{sen}$ values estimated experimentally and theoretically. The coincidence is indicative. These results indicate there is an increase in intermittency and in multifractality and are in accordance with the results presented previously corresponding to the experimental estimation of intermittency and multifractality of the dynamics.

**Table 1**

| Timeseries | $\mu$ * | $\alpha_0$ | X * | $q_{sen}$ * Theoretical | $q_{sen}$ Experimental |
|---|---|---|---|---|---|
| X1 Calm | 0.096 | 1.054±0.001 | 0.110 | -0.251±0.027 | -0.2422 ± 0.00009 |
| X2 Calm | 0.18 | 1.103±0.001 | 0.212 | 0.224±0.009 | 0.2713± 0.0024 |
| X3 Calm | 0.163 | 1.093±0.001 | 0.191 | 0.160±0.011 | 0.0904 ± 0.0022 |
| X1 Shock | 0.177 | 1.101±0.001 | 0.208 | 0.214±0.010 | 0.2731 ± 0.0455 |
| X2 Shock | 0.255 | 1.147±0.001 | 0.303 | 0.422±0.005 | -0.0982 ±0.0031 |

**Table 1:** Theoretically estimated values ($\mu$, X, $q_{sen}$) versus the experimentally estimated values ($q_{sen}$) for the Solar Wind time series, calm (x1, x2, x3) and shock period (x1,x2). (* Correspond to theoretically estimated values using Tsallis theory).

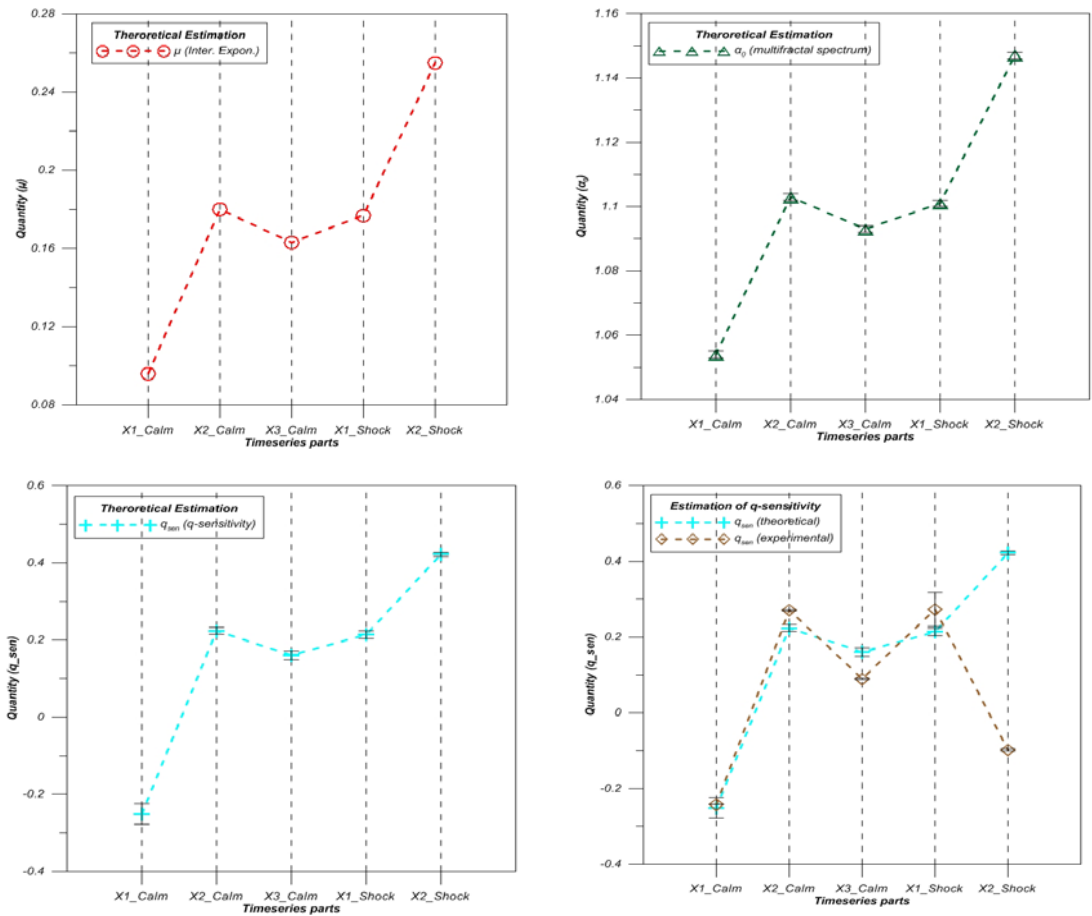

**Figure 8.a:** Theoretically estimated values ($\mu$ – intermittency exponent) for the solar wind time series, calm (x1, x2, x3) and shock period (x1,x2). (The theoretically values correspond to theoretically estimated values of Tsallis theory).**b:** Theoretically estimated values ($\alpha_0$) for the solar wind time series, calm (x1, x2, x3) and shock period (x1,x2). (The theoretically values correspond to theoretically estimated values of Tsallis theory).**c:** Theoretically estimated values ($q_{sen}$) for the solar wind time series, calm (x1, x2, x3) and shock period (x1,x2). (The theoretically values correspond to theoretically estimated values of Tsallis theory).**d:** Theoretically and experimentally estimated values ($q_{sen}$) for the solar wind time series, calm (x1, x2, x3) and shock period (x1,x2). (The theoretically values correspond to theoretically estimated values of Tsallis theory).

### 3.6.2 Determination of the $q_{stat}$

In Figure 9a we present (by open circles) the logarithm of the experimental probability distribution function (PDF) $\ln_q[p(z)]$ vs. $z$, where $z$ corresponds to the $z_{n+1} - z_n$, $(n = 1, 2, ..., N)$ ion flux solar wind time series values for the first calm period segment, x1calm. In Fig. 9b we present the best linear correlation between $\ln_q[p(z)]$ and $z^2$. The best fitting was found for the value of $q_{stat} = 1.37 \pm 0.05$. This value was used to estimate the $q$-Gaussian distribution presented in Fig. 9a by the solid green line, while the blue line corresponds to the Gaussian distribution. Figure 9[c,d] are similar to Fig. 9[a,b] but for the first shock period segment, x1shock. The $q_{stat}$ values were found to be: $q_{stat} = 1.61 \pm 0.01$. According to these results the following relation is satisfied: $1 < q_{stat}(calm) < q_{stat}(shock)$, indicating that there is a significant difference between the calm and the shock period as far as the nonequilibrium metastable solar wind dynamics is concerned. This difference is clearly depicted in the increase of the index $q_{stat}$ indicating the strengthening of the non-Gaussian and Tsallis non-extensive profile of the underlying dynamics.

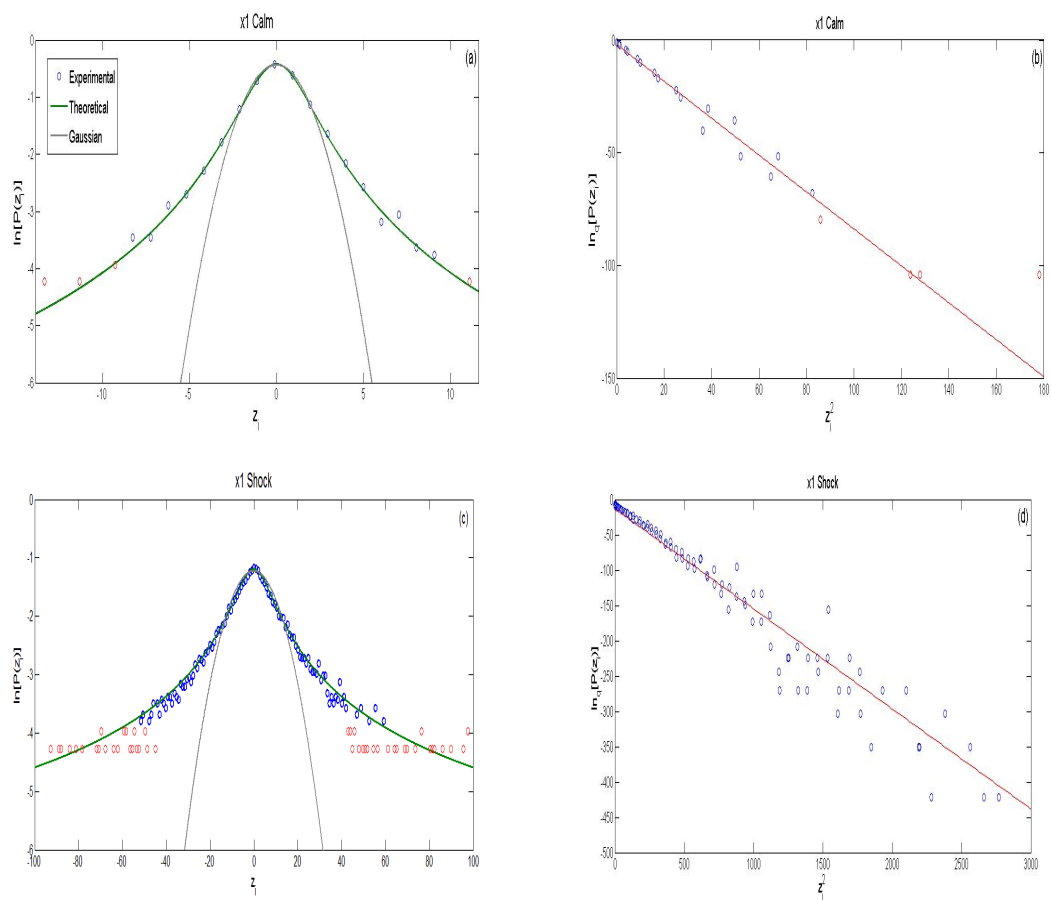

**Figure 9**: (a) PDF $\ln_q[p(z)]$ vs. $z_i$ q Guassian function that fits P($z_i$) for the Solar Wind calm period time series (b) Linear Correlation between $\ln_q p(z_i)$ and $(z_i)^2$ where $q = 1.37 \pm 0.05$ for the Solar Wind

calc period time series **(c)** PDF P($z_i$) vs. $z_i$ q Gaussian function that fits P($z_i$) for the Solar Wind shock period time series **(d)** Linear Correlation between $\ln_q p(z_i)$ and $(z_i)^2$ where $q = 1.61 \pm 0.01$ for Solar Wind shock period time series.

Furthermore, figure 10 presents the variation of the index $q_{stat}$, along with the bar errors, as the solar wind dynamics evolves towards the shock event and its relaxation. As it can be seen in this figure there is a continuous increase in $q_{stat}$ values, with the maximum value corresponding to the main shock period, namely x1shock segment. This crescendo corresponds to the gradual development of non-Gaussian, non-extensive solar wind dynamics, which reach its peak in the main shock event (x1shock). Afterwards, there is a significant decrease in $q_{stat}$ value, a result that indicates the relaxation of the phenomenon towards to another metastable stationary state. Finally, it must be noted that the relative big values of uncertainty ($\Delta q_{stat}$) can be attributed to the fluctuations of the fractal structure of the underlying dynamics (fracton theory).

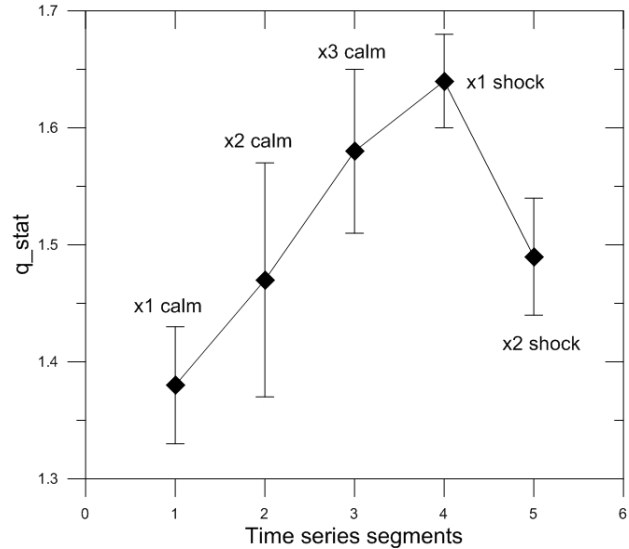

**Figure 10**: The index $q_{stat}$ for five ion flux time series segments of the solar wind.

### 3.6.3 Determination of the $q_{rel}$

In figure 11 we present the best $\ln_q I(\tau)$ fittings of the mutual information functions $I(\tau)$ of the five ion flux time series segments estimated for the calm and shock periods. The results showed that the $q_{rel}$ index was found to be $q_{rel}$(quiet) = 8.158 ± 0.117 for the first segment of quiet period, x1calm, and $q_{rel}$(shock)= 9.772 ± 0.2947 for the first segment of the shock active period, x1shock. The $q_{rel}$ increases passing from the quiet to the shock period a result that reveals strengthening of the long range correlations and non-extensive character. However, there is also a big fluctuation of $q_{rel}$ index concerning all the time period revealing strong far equilibrium fluctuations that dominate the system before and after the main shock. This result can also be attributed to the fluctuations of the fractal structure of the underlying dynamics.

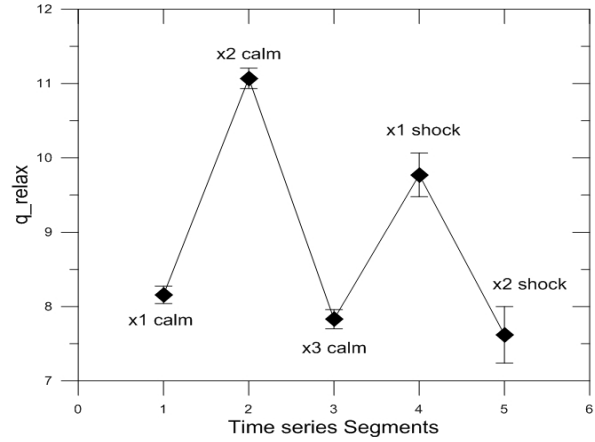

**Figure 11**: The index $q_{relax}$ for the five ion flux time series segments of the solar wind.

### 3.7 Summarizing results of Tsallis statistics concerning 3 different shock events

In this section we present summarizing results concerning all three shock events. Continuously, we provide information about the two shock events, (A) and (B), shown in Fig. 12: The example «A» corresponds to the shock of 24 October 2011. The shock occurred at ~ 18.35 UT and the at the moment of the shock front the ion flux grow from 4 up to $20*10^8$ $cm^{-2}$ $s^{-1}$. The changing of solar wind parameters on shock fronts are the following value: density from 10 to 35 $cm^{-3}$, velocity from 380 to 500 km/s, temperature from 5 to 30 eV. The example «B» corresponds to the shock of 09 September 2011. The shock occurred at ~ 12.40 UT. At the moment of the shock, ion flux grow from 10 up to $30*10^8$ $cm^{-2}$ $s^{-1}$, and after the front ion flux continue to grow up to $50*10^8$ $cm^{-2}$ $s^{-1}$, and in 13.15 decrease rather fast up to $5*10^8$ $cm^{-2}$ $s^{-1}$. The changing of solar wind parameters during this event are the following value: density grows on the shock front from 15 to 65 $cm^{-3}$ and after grows up to 110 $cm^{-3}$ and in 13.15 UT decrease up to 10 $cm^{-3}$ , velocity grow on the shock front from 360 to 410 km/s after grow slowly up to 450 km/s, temperature practically doesn't change on the shock front and after grow slowly up to 15 eV.

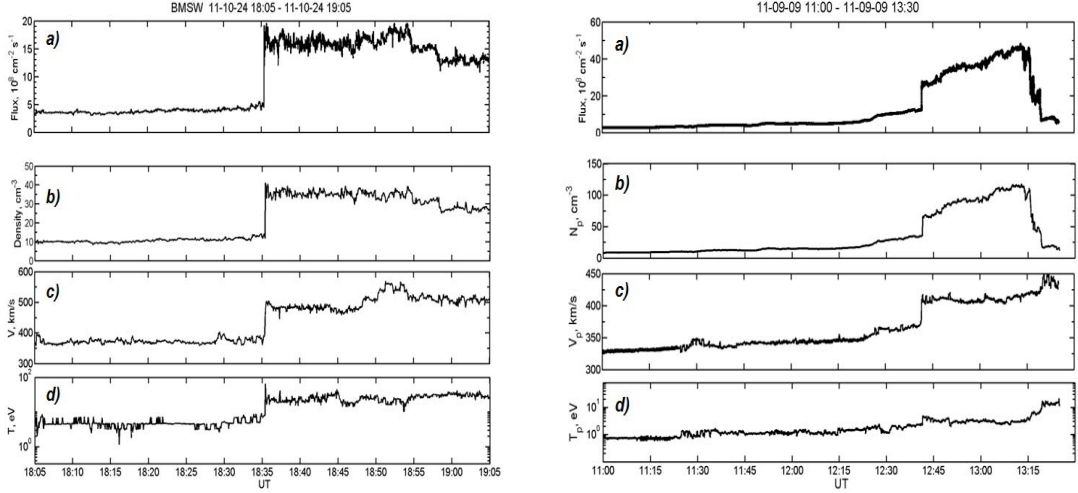

**Figure 12**: The Solar wind parameters Ion Flux (a), density (b), velocity (c), temperature (d) on shock front 24.10.2011 (left panel) and 09.09.2011 (right panel).

In Table 2 we summarize the results concerning the estimation of the Tsallis $q$-triplet ($q_{rel}$, $q_{stat}$, $q_{sen}$), the multifractal range $\Delta(a)$ and the generalized dimension spectrum range $\Delta(D_q)$, as well as the normalized Tsallis entropy, for the three shock events, both for calm and shock period. As far as the Tsallis $q$-triplet is concerned, the results showed clearly a difference in the dynamics from calm to shock period. In particular, for the shock event (A) we found: $q_{calm}^{sens} > q_{shock}^{sens}$, $q_{calm}^{stat} < q_{shock}^{stat}$, $q_{calm}^{relax} < q_{shock}^{relax}$, for shock event (B) $q_{calm}^{sens} < q_{shock}^{sens}$, $q_{calm}^{stat} < q_{shock}^{stat}$, $q_{calm}^{relax} > q_{shock}^{relax}$, while as we mentioned in previous paragraphs for shock event (C) $q_{calm}^{sens} < q_{shock}^{sens}$, $q_{calm}^{stat} < q_{shock}^{stat}$, $q_{calm}^{relax} < q_{shock}^{relax}$. Moreover, as far as the multifractal and generalized dimension spectrum range is concerned, the results revealed a decrease from calm to shock period ($\Delta a = 1.0106 \rightarrow 0.9061$ and $\Delta(D_q) = 0.9493 \rightarrow 0.8861$) for shock event (A), while there is a significant increase for (B) event ($\Delta a = 0.7023 \rightarrow 0.9707$ and $\Delta(D_q) = 0.6661 \rightarrow 0.9705$) and (C) event ($\Delta a = 0.7478 \rightarrow 1.2045$ and $\Delta(D_q) = 0.6981 \rightarrow 1.2106$). Finally, the normalized Tsallis entropy shows an increase for (A) shock event ($S_q = 15.47 \rightarrow 20.62$), while for the other two events (B) ($S_q = 199.8 \rightarrow 48.05$) and (C) ($S_q = 111 \rightarrow 20$) a significant decrease of five order less, in both cases.

Thus, the aforementioned results indicate differences in the underlying dynamics of the three shocks and in the way they are manifested through ion flux time series. These differences are depicted in $q_{sen}$ and $q_{rel}$ indices, in multifractal and generalized dimension spectrum range and in normalized Tsallis entropy. On the contrary, the $q_{stat}$ index showed the same variation (increase) as the transition from calm to shock period takes place. However, further investigation is needed in order to reveal a clearer picture in the differences and similarities in the dynamics of the three shock events, as far as the Tsallis statistics is concerned.

| Table 2. | | | | | | |
|---|---|---|---|---|---|---|
| **Indices** | **Shock Event** | | | | | |
| | **A Calm** | **A Shock** | **B Calm** | **B Shock** | **C Calm** | **C Shock** |
| $q_{rel}$ | 3.12 | 3.57 | 12.89 | 10.78 | 8.158 | 9.772 |
| $q_{stat}$ | 1.37 | 1.52 | 1.41 | 1.78 | 1.38 | 1.64 |
| $q_{sen}$ | 0.0796 | 0.0217 | -0.4384 | -0.0071 | -0.2422 | 0.2731 |
| $\Delta\alpha$ | 1.0106 | 0.9061 | 0.7023 | 0.9707 | 0.7478 | 1.2045 |
| $\Delta(D_q)$ | 0.9493 | 0.8861 | 0.6661 | 0.9705 | 0.6981 | 1.2106 |
| Entropy | 15.47 | 20.62 | 199.8 | 48.05 | 111 | 20 |

***Table 2:*** *Summarized parameter values of solar wind dynamics of three different shock phenomena including the calm and shock period: From the top to the bottom we show: The q-triplet ($q_{sen}, q_{stat}, q_{rel}$) of Tsallis. Changes of the ranges Δα of the multifractal profile. The difference the Dimension Spectra Dq. The Tsallis entropy.*

## 4. Summary of data analysis results and discussion

As far as the shock event (C), we have observed the following phenomenology:

- The shock event starts with forerunner or precursory phenomena observed $\sim 90$ minutes before the main shock event.

- There is a gradual enhancement of ion flux oscillations as the shock event is developed.

- Clear non- extensive statistical character of solar wind was observed during the calm or shock periods.

- Gradual enhancement of the non-Gaussian character, as the Flatness coefficient *F* clearly increases to values much higher than 3.

- Strong reduction of Tsallis entropy production was observed 60-90 minutes before the main shock event.

- The nonextensive character of statistics was found to be gradually enhanced passing from the calm period to the shock state, as the index $q_{stat}$ increases from values lower than 1.4 to values higher than 1.6.

- At the shock event, the Tsallis *q*-triplet parameters of the non-extensive statistics change drastically according to the following general scheme: $q_{calm}^{sens} < q_{shock}^{sens}$, $q_{calm}^{stat} < q_{shock}^{stat}$, $q_{calm}^{relax} < q_{shock}^{relax}$.

- The multifractal character is strengthened passing from the calm to the shock period, as it is concluded by the profile of singularity spectrum $f(\alpha)$ and the width $\Delta a = (a_+ - a_-)$ variation from the calm to the shock state ($\Delta a_{calm} < \Delta a_{shock}$).

- The intermittent exponent ($\mu$) also increase passing from the calm to shock state.

- The parameter $p$ of the $p$-model estimated from the nonlinear best fitting of the $D_{\tilde{q}}$ data was found to increase passing from calm to shock period.

- Faithfull coincidence can be observed between the experimentally estimated singularity spectrum $f(\alpha)$ and the $q_{sen}$ parameter values and the correspondent values estimated by using the Tsallis $q$-entropy principle.

- Fluctuations of all parameters were found, a result which indicates the presence of fracton dynamics.

- Significant coincidence of the experimental results with the theoretical estimations based on Tsallis theory, concerning the transition from calm to shock period.

As far as the comparison of the three shock events the following were found:

- Differences in Tsallis $q_{rel}$ and $q_{sen}$ indices between the three events, as the transition from calm to shock period takes place. In particular, for shock (A) and (C) the $q_{rel}$ index is increased as the transition takes place, while for shock (B) is decreased. Furthermore, for shock (A) the $q_{sen}$ index is decreased, while for (B) and (C) is increased.

- For all the shock events, the $q_{stat}$ index showed the same variation (increase) as the transition from calm to shock period takes place.

- Differences were also found concerning the multifractal and generalized dimension spectrum range and in normalized Tsallis entropy.

The phenomenology of solar wind shock events, as it was summarized previously, reveals clearly the existence of a dynamical non-equilibrium phase transition process related to the solar wind shock event. This non-equilibrium phase transition process includes the transition of the original wind complex calm state to states which include enhancement of self organization and intermittency. Although there are strong theoretical arguments suggesting that the solar wind phase transition process and more generally the non-linear turbulence solar wind processes originate on the solar surface (Milovanov & Zelenyi 1992, 1993) certainly the solar wind process plasma exhibits internally dynamical effects related to the observed transformation of the non-extensivity character and the enhancement of the self organization process.

Solar wind is a permanent plasma flow of solar origin that fills the solar system formed in the supersonic expansion of the solar corona into the interplanetary space. The expansions dynamics depend on a number of factors including the magnetic field topology in the corona. The solar wind coming from regions with predominantly radial magnetic field components is accelerated more uniformly and quietly, reaching velocities about $\approx 600\text{-}700$ km s$^{-1}$. On the contrary, the solar wind plasma flow that touches the tangential magnetic field is more irregular and its velocities are $\approx 350$ km s$^{-1}$. The fast solar wind is formed over the coronal holes, while the slow one is formed above the regions with complex field configuration. Within 1 Au the interplanetary medium can conserve the structural characteristics of the low corona where the solar wind flow is born. During the solar plasma expansion the solar wind develops strong turbulent character. In this framework the shock events include both types of internal complex dynamics which can be underlying to the non-equilibrium phase transition processes, as well as phase transition processes related to the coronal regions where the plasma flow is born.

Moreover, as we show in the next section, the solar wind plasma system can include fracton excitations and fracton dynamics where fracton formations are waves on fractal structures. Fracton dynamics can cause the oscillations of statistical parameters observed during shock events development as both types of solar wind phenomena, fracton oscillations and non-equilibrium phase transition processes, can be related to plasma tube – tube interactions.

 The solar wind clusters and magnetized plasma tubes are formed at the solar corona base and across the solar surface, while their interaction in the interplanetary space can cause the changes in the internal solar wind fractional dynamics presented in the next sections.

The results described previously are also in agreement with the general theory of complex plasma dynamics. According to Zelenyi and Milovanov (2004) the complex character of the solar wind plasma can be described as non-equilibrium (quasi)-stationary states (NESS) having the topology of a percolating fractal set. These scales include multi-scale interactions of fields and particles (currents) and can be related to the simultaneous development of numerous instabilities interfering with each other. The plasma complex state corresponds to the stabilization near the turbulent NESS identified with the generalized symmetries of a fractal disk diffeomorphism to a fractal set at the percolation threshold. The structural stability of the NESS as a symmetric turbulent phase is maintained due to multi-scale correlations creating the existence of local extremes of the free energy.

In addition, the interpretation of the results indicate the possibility for the existence of solar wind plasma system phase transitions from a weak NESS to a strong NESS as the outcome of cluster interaction in the interplanetary space, as well as at the solar corona source of solar wind. In the next section, we present modern theoretical concepts which constitute the appropriate theoretical framework for the interpretation of the results of this study concerning the complexity of the solar wind plasma.

## 5. Theoretical Interpretations of Data Analysis Results

Solar Wind plasma is a typical case of stochastic spatiotemporal distribution of physical magnitudes such as force fields (B, E fields) and matter fields (particle and current densities or bulk plasma distributions). The classical MHD description or the Boltzmann –Maxwell statistical mechanics are inefficient to describe the non-equilibrium solar wind state as they include smooth and differentiable spatial-temporal functions (MHD theory) or Gaussian statistical processes (Boltzmann-Maxwell statistical mechanics), correspondingly.

The results of this study, can be better understood in the framework of modern theoretical concepts concerning non-extensive statistical mechanics (Tsallis,2009), fractal topology (Zelenyi and Milovanov, 2004), turbulence theory (Frisch,1996), strange dynamcics (Zaslavsky, 2002), percolation theory (Milovanov, 1997), anomalous diffusion theory and anomalous transport theory (Milovanov, 2001), fractional dynamics (Tarasov, 2007) and non-equilibrium phase transition theory (Chang, 1992).

### 5.1 The $q$-extension of Central Limit Theorem and the $q$-triplet of Tsallis

The results showed clearly the non-extensive character of the solar wind plasma and the multi-scale strong correlations from the microscopic to the macroscopic level indicating the inefficiency of classical MHD or plasma statistical theories, based on the classical central limit theorem, to explain the complexity of the solar wind dynamics. According to the classical central limit theorem the probability density functions of a sum of independent random variables are Gaussian when the single variables have long range tails. However, the non-Gaussian with heavy tails probability distribution functions, which were observed in the solar wind plasma statistics, are related to the $q$-extension of central limit theorem.

Tsallis non-extensive statistical mechanics includes the $q$-generalization of the classic central limit theorem (CLT) as a $q$-generalization of the Levy – Gnedenko central limit theorem (Umatov et al., 2008) applied for globally correlated random variables. The $q$-generalization of CLT based at the $q$-Fourier transform of a $q$-Gaussian can produce an infinite sequence $(q_n)$ of $q$-parameters by using the function $Z(s) = \dfrac{1+s}{3-s}$, $s \in (-\infty, 3)$ and its inverse $z^{-1}(t), t \in (-1, \infty)$. It can be shown that $z(1/z(s)) = 1/s$ and $z(1/s) = 1/z^{-1}(s)$. Then is $q_1 = z(q)$ and $q_{-1} = z^{-1}(q)$ it follows that:

$z\left(\dfrac{1}{q_1}\right) = \dfrac{1}{q}, z\left(\dfrac{1}{q}\right) = \dfrac{1}{q-1}$ and $q_{-1} + \dfrac{1}{q_1} = 2$. The set of all $q$-Gaussians $G_q(\beta, x)$ be denoted by:

$$\Upsilon_q = \left\{ b\, G_q\left(\beta, x\right) : b > 0, \beta > 0 \right\} \qquad (76)$$

For $q$-Gaussians the $q$-Fourier transform hold as follows:

$$F_q\left[ G_q\left(\beta, x\right) \right](\xi) = e_{q_1}^{-\beta_x(q)\xi^2} \qquad (77)$$

where $q_1 = z(q), 1 \le q < 3$.

The $q$-Fourier transform is defined by Umarov, Tsallis and Steinberg (Umarov et.al., 2007) by the formula:

$$F_q[f](\xi) = \int e_q^{ix\xi} \otimes_q f(x) dx \qquad (78)$$

For the inverse $q$-Fourier transform we have the following formula:

$$F_{q-1}\left[ G_{q-1}\left(\beta; x\right) \right](\xi) = e_q^{-\beta_x(q-1)\xi^2} \qquad (79)$$

where $q - 1 = z^{-1}(q), 1 \le q < 3$ and $\beta_x(s) = \dfrac{3-s}{8\beta^{2-s} C_s^{2(s-1)}}$.

The function $z(s)$ permits the following $q$-Fourier transform mappings:

$$F_q : \mathfrak{I}_q \to \mathfrak{I}_{q_1}, q_1 = z(q), 1 \le q < 3 \qquad (80)$$

$$F_{q-1} : \mathfrak{I}_{q-1} \to \mathfrak{I}_q, q_{-1} = z^{-1}(q), 1 \le q < 3 \qquad (81)$$

as well as the inverse mappings by using the inverse $q$-Fourier transforms:

$$F_q^{-1} : \mathfrak{I}_{q_1} \to \mathfrak{I}_q, q_1 = z(q) \qquad (82)$$

$$F_{q-1}^{-1} : \mathfrak{I}_q \to \mathfrak{I}_{q-1}, q_{-1} = z^{-1}(q), 1 \le q < 3 \qquad (83)$$

We can introduce the sequence $q_n = z_n(q) = z(z_{n-1}(q)), n = 1, 2, \ldots$ starting from the initial $q = z_0(q), q < 3$. The $q_n$ sequence can be extended for negative integers n=-1,-2,... by $q_{-n} = z_{-n}(q) = z^{-1}(z_{1-n}(q))$. The sequence $q_n$ can be given through the following mappings:

$$\xrightarrow{z} q_{-2} \xrightarrow{z} q_{-1} \xrightarrow{z} q_0 = q \xrightarrow{z} q_1 \xrightarrow{z} q_2 \xrightarrow{z} \ldots$$

$$\longleftarrow \overset{z^{-1}}{} q_{-2} \longleftarrow \overset{z^{-1}}{} q_{-1} \longleftarrow \overset{z^{-1}}{} q_0 = q \longleftarrow \overset{z^{-1}}{} q_1 \longleftarrow \overset{z^{-1}}{} q_2 \longleftarrow \overset{z^{-1}}{} \dots$$

while the duality relations hold:

$$q_{n-1} + \frac{1}{q_{n+1}} = z, n \in \mathbb{Z} \qquad (84)$$

The q-generalization of the central limit theorem consistent with non-extensive statistical mechanics is as follows:

For a sequence $q_k, k \in \mathbb{Z}$ with $q_k \in [1,2]$ and a sequence $x_1, x_2, x_N, \dots$ of $q_k$-independent and identically distributed random variable then the $z_n = x_1 + x_2 + x_N + \dots$ is also a $q_{k-1}$-normal distribution as $N \to \infty$, with corresponding statistical attractor $G_{q_{k-1}}(\beta_{k_j} x)$.

The q-independence corresponds to the relations:

$$F_q(x+y)(\xi) = F_q[x](\xi) \otimes_q F_q[y](\xi) \qquad (85)$$

$$F_{q-1}(x+y)(\xi) = F_{q-1}[x](\xi) \otimes_q F_q[y](\xi) \qquad (86)$$

where $q = z(q_{-1})$. The q-independence means independence for $q = 1$ and strong correlation for $q \neq 1$ (Tsallis 2004).

The $q$-CLT states that an appropriately scaled limit of sums of $q_k$ correlated random variables is a $q_k$ – Gaussian, which is the $q_k^*$-Fourier image of a $q_k^*$- Gaussian. The $q_k$, $q_k^*$ are sequences:

$$q_k = \frac{2q + k(1-q)}{2 + k(1-q)} \text{ and } q_k^* = q_{k-1} \text{ for } k = 0, \pm 1, \pm 2, \dots \qquad (87)$$

including the triplet ($P_{att}$, $P_{cor}$, $P_{scl}$), where $P_{att}$, $P_{cor}$ and $P_{scl}$ are parameters of attractor, correlation and scaling rate respectively and corresponds to the $q$-triplet ($q_{sens}$, $q_{rel}$, $q_{stat}$) according to the relations (Umarov et al., 2008) :

$$(P_{att}, P_{cor}, P_{scl}) \equiv (q_{k-1}, q_k, q_{k+1}) \equiv (q_{sens}, q_{rel}, q_{stat}) \qquad (88)$$

The parameter $P_{att} \equiv q_{sens} \equiv q_{k-1}$ describes the non-ergodic $q$-entropy production of the multiscale correlated process as the system shifts to the state of the $q_{att}$ – Gaussian, where the $q$-entropy is extremized in accordance with the generalization of the Pesin's theorem (Tsallis, 2004c):

$$K_{qsen} \equiv \lim_{t \to \infty} \lim_{W \to \infty} \lim_{M \to \infty} \frac{S_q(P_i(t)) / k}{t} = \lambda_{qsen} \qquad (89)$$

The parameter $P_{cor} \equiv q_{rel} \equiv q_k$ describes the $q$-correlated random variables participating to the dynamical process of the $q$-entropy production and the relaxation process toward the stationary state.

The parameter $P_{slc} \equiv q_{stat} \equiv q_{k+1}$ describes the scale invariance profile of the stationary state corresponding to the scale invariant $q$-Gaussian attractor as well as to an anomalous diffusion process mirrored at the variance scaling according to general, asymptotically scaling, from:

$$N^D P_x(x) \sim G(\frac{x}{N^D}) \qquad (90)$$

Where $P_x(x)$ is the probability function of the self similar statistical attractor $G$ and $D$ is the scaling exponent characterizing the anomalous diffusion process (Baldoving-Stella, 2007):

$$\langle x^2 \rangle \sim t^{2D} \qquad (91)$$

The non-Gaussian multi-scale correlation can create the intermittent multi-fractal structure of the phase space mirrored also in the physical space multi-fractal distribution of the turbulent dissipation field. The multi-scale interaction at non-equilibrium critical NESS creates the heavy tail and power law probability distribution function obeying the $q$-entropy principle. The singularity spectrum of a critical NESS corresponds to extremized Tsallis $q$-entropy.

In this framework of theoretical modeling of distributed random fields, the fractal – multifractal character of the solar wind system at every NESS (calm or shock) as it was shown in this study, indicate the existence of physical local singularities for the spatial-temporal distribution of solar wind variables. The singularity behavior of a given random field function $f(x)$ at a given point $x_*$ is defined as the greatest exponent $h$ so that $f$ is Lipschitz $h$ at $x_*$ (Arnedo et al., 1996). The Hölder exponent $h(x_*)$ measures how irregular $f$ is at the point $X_*$ according to the relation:

$$\left| f(x) - P(x - x_*) \right| \leq C \left| x - x_* \right|^h \qquad (92)$$

The smaller the exponent $h(x_*)$ is the more irregular (singular) is the function $f$ at the point $x_*$. On a self similar fractal set the singularity strength of the measure $\mu$ at the point $x$ is described by the scaling relation:

$$\mu(B_x(\varepsilon)) \simeq \int_{B_x(\varepsilon)} d\mu(y) \sim \varepsilon^{a(x)} \qquad (93)$$

Where $B_x(\varepsilon)$ is a ball of size $(\varepsilon)$ centered at $x$ and $\mu(B_x)$ is the fractal mass in the $B_x$ region. Homogeneous measures are characterized by a singularity spectrum supported by a single point $(a_0, f(a_0))$ while multifractal sets involve singularities of different strengths $(\alpha)$ described by the singularity spectrum $f(\alpha)$.

The singularity spectrum is defined as the Hausdorff dimension of the set of all points x such that $a(x) = a$. The singularity strength $a(x)$ of the fractal mass measure is related to the local fractal dimension $D$ of the fractal set. More generally, the singularity spectrum associated with singularities $h$ (Hölder exponents) of a distributed random variable is given follows:

$$D(h) = d_H\left\{x : h(x) = h\right\} \qquad (94)$$

That is $D(h)$ is the Hausdorff dimension of the set of all points x such as $h(x)=h$. For the solar wind system the singularity spectrum $D(h)$ corresponds to the scaling invariance of the MHD description according to the scaling transformation of the MHD equations:

$$r \rightarrow r'\lambda^{-1}, u_r \rightarrow u_r.\lambda^{-h}, b_r \rightarrow b_r.\lambda^{-h} \qquad (95)$$

where $h$ is a free parameter and $u_r, b_r$ are the velocity and magnetic fields of the solar wind plasma system, correspondingly (Carbone et al., 1996). The $a(x)$ singularity exponent of the fractal mass measure and the $f(\alpha)$ singularity spectrum describes the scaling of the energy dissipation in the solar wind turbulent system.

The fractal-multifractal structure of the solar wind plasma system indicates the generalization of the classical field-particle dynamics to the fractional dynamics of the solar wind system, since the functions of the distributed physical system's variables are irregular and they are produced by fractional dynamics on fractal structures. The differentiable nature of smooth distribution of the macroscopic picture of physical processes is a natural consequence of the Gaussian microscopic randomness which, through the classical CLT, is transformed to the macroscopic, smooth and differentiable processes. The classical CLT is related to the condition of time-scale separation, where at the long-time limit the memory of the microscopic non-differentiable character is lost. On the other hand, the $q$-extension of CLT induces the nonexistence of time-scale separation between microscopic and macroscopic scales as the result of multiscale global correlations which produce fractional dynamics and singular functions of spatio-temporal dynamical physical variables.

## 5.2 Fractional Calculus

Fractal sets are measurable metric sets with non-integer Hausdorff dimension. The elements of a fractal set can be represented by $n$-tuples of real numbers $x = (x_1, x_2, ..., x_n)$ such that a fractal set $F$ is embedded in $R^n$. A fractal function is defined on a fractal set as follows:

$$f(x) = \sum_{i=1}^{\infty} \beta_i X_{E_i}(x) \qquad (96)$$

where $X_E$ is the characteristic function of $E$. The continuous function $f(x)$ is defined as follows:

$\lim_{x \to y} f(x) = f(y)$ whenever $d(x, y) = 0$ for the metric $d(x, y)$ defined in $R^n$ and for points $(x, y) \in f$. The Hausdorff measure $\mu_H$ of a subset $E \in F$ is defined by:

$$\mu_H(E, D) = W(D) \lim_{d(E_i) \to 0} \inf_{\{E_i\}} \sum_{i=1}^{\infty} |d(E_i)|^D \qquad (97)$$

Where $E \subset U_{i=1}^{\infty} E_i$, D is the Hausdorff dimension of $E \subset U$, $d(E_i)$ are the diameters of $\{E_i\}$ and $W(D)$ for balls $\{E_i\}$ covering $F$ is given by

$$W(D) = \frac{\Pi^{D/2} 2^{-D}}{\Gamma\left(D/2 + 1\right)} \qquad (98)$$

The Lebesque – Stieltjes integral over a $D$-dimensional fractal set of a function $f(x)$ is defined by

$$\int_F f d\mu \equiv \sum_{i=1}^{\infty} \beta_i \mu_H(E_i) \qquad (99)$$

And it can be proved to be given by the relation:

$$\int_F f d\mu_H = \frac{2\Pi^{D/2} \Gamma(D)}{\Gamma\left(D/2\right)} \left(I_o^D f\right) \qquad (100)$$

Where $\left(_z I_x^a f\right)(z) \equiv \frac{1}{\Gamma(a)} \int_z^{\infty} (x - x')^{a-1} f(x') dx'$

is the Riemann Liouville fractional integral (Tarassov, 2007; Mainardi, 2010). The last relation connects the integral on fractals with fractional integrals and permits the application of different tools of the fractional calculus for the fractal medium. Respectively to the Riemann Liouville fractional integral on a fractal set $F$ we can define the Riemann Liouville fractional derivative by

$$_zD_x^a f(x) = \frac{1}{\Gamma(n-a)} \frac{\partial^n}{\partial x^n} \int_z^x \frac{f(x')}{(x-x')^{a-n+1}} dx', \ n-1 < a < n \tag{101}$$

The nonlocal character is evident in both cases of fractional derivative and integral on a fractal set. The nonlocal character of fractional calculus is related to multiscale and self-similar character of the fractal structure. The fractional extension of integral and differential calculus can be used for the description of the non-local multiscale phenomena described by Maxwell's equations, the Magnetohydrodynamics of fractal plasma states, or the Fokker-Planck equation of fractal media (Tarassov, 2005, 2007, 2008). The solution of the fractional equations correspond to fractional non-differentiable singular self-similar functions as we can observe at the experimental data.

Milovanov and Zelenyi (1999) introduced the fractional wave equation for the description of solar wind fracton excitations, namely fractional waves on fractals. The fractional extension of wave equations for spherically symmetric vibrations of fractals includes temporal and spatial fractional derivatives as follows:

$$\boldsymbol{D}_t^\gamma \psi(t, \dot{x}) = \frac{1}{x^{D-1}} \frac{\partial}{\partial x}\left( x^{D-1} x^{-\frac{\partial \psi}{\partial t}} \right) \tag{102}$$

Where

$$\boldsymbol{D}_t^\gamma \psi(t,x) \equiv \frac{\partial^\gamma \psi}{\partial t^\gamma} = \Gamma^{-1}(m-\gamma) \frac{\partial^m}{\partial t^m} \int_0^t \frac{\psi(\tau,x)}{(t-\tau)^{1+\gamma-m}} d\tau \tag{103}$$

is the fractional time derivative, $D$ is the fractal dimension of space and $\theta$ is the connectivity index of the fractal space (see next section 5.4). The nonlocal character of the fractional wave equation on fractals is related to the correlated and coherent character of local dynamical events on a self-similar structure. The solution of the fractional wave equations contains exponentially decaying factors which are absent in Euclidian solutions. These factors are responsible for the spatial localization of fractons, while the strength of localization is described by the connectivity index θ, through the localization exponent $2 + \theta/2$, which coincides with the Hausdorff dimension of the "geodesic lines" on the fractal space. The localization of waves on fractals (fractons) can be used for the explanation of local variation of solar wind characteristics, especially during shock events. More generally, the basic equations characterizing space plasmas (Chang 1992; 2002) such as:

$$\begin{aligned} \rho \, d\mathbf{V}/dt &= \mathbf{B}\nabla\mathbf{B} + ...., \\ d\mathbf{B}/dt &= \mathbf{B}\nabla\mathbf{V} + .... \end{aligned} \tag{104}$$

where (….) are the magnetic – bulk plasma flow fields, or any other kind of dynamical description of plasmas must be extended to fractional equations correspondingly, when the (multi)-fractal and intermittent turbulent character is evident.

## 5.3 Anomalous diffusion and strange dynamics

Nonlinear dynamics can create fractal structuring of the phase space and global correlations in the nonlinear system. For nonextensive systems the entire phase space is dynamically not entirely occupied (the system is not ergodic), but only a scale-free –like part of it is visited yielding a long-standing (multi)-fractal-like occupation. According to Milovanov and Zelenyi (2000), Tsallis entropy can be rigorously obtained as the solution of a nonlinear functional equation referred to the spatial entropies of the subsystems involved including two principal parts. The first part is linear (additive) and leads to the extensive Boltzman-Gibbs entropy. The second part is multiplicative corresponding to the nonextensive Tsallis entropy referred to the long range correlations. The fractal –multifractal structuring of the phase space makes the effective number $W_{\text{eff}}$ of possible states, namely those whose probability is nonzero, to be smaller ($W_{\text{eff}} < W$) than the total number of states.

According to Zaslavsky (2002) the topological structure of phase space of nonlinear dynamics can be highly complicated including trapping and flights of the dynamics through a self-similar structure of islands. The island boundary is sticky making the dynamics to be locally trapped and "stickiness". The set of islands is enclosed within the infinite fractal set of cantori causing the complementary features of trapping and flight being the essence of strange kinetics and anomalous diffusion.

The dynamics in the topologically anomalous phase space corresponds to a random walk process which is scale invariant in spatial and temporal self-similarity transform:

$$\hat{R} : t' \to \lambda_t t, \xi' = \lambda_\xi \xi \qquad (105)$$

The spatial-temporal scale invariance causes strong spatial and temporal correlations mirrored in singular self-similar temporal and spatial distribution functions which satisfy the fractional generalization of classical Fokker-Planck-Kolmogorov equation (FFPK – equation) (Zaslavsky, 2000):

$$\frac{\partial^\beta P}{\partial t^\beta} = \frac{\partial^\alpha}{\partial (-\xi)^\alpha} (AP) + \frac{1}{2} \frac{\partial^{2\alpha}}{\partial (-\xi)^{2\alpha}} (BP) \qquad (106)$$

Where $P \equiv P(\xi, t)$ is the probability density of the state ($\xi$) at the time ($t$). The critical components ($\alpha, \beta$) correspond to the fractal dimension of spatiotemporal nongaussian distribution of the temporal-spatial functions-processes or probability distributions. The quantities A, B are given by

$$A = \lim_{\Delta t \to 0} \frac{\left\langle\left\langle \left| \Delta \xi^{\alpha} \right| \right\rangle\right\rangle}{(\Delta t)^{\beta}}, B = \lim_{\Delta t \to 0} \frac{\left\langle\left\langle \left| \Delta \xi^{2\alpha} \right| \right\rangle\right\rangle}{(\Delta t)^{\beta}} \qquad (107)$$

where <<...>> denotes a generalized convolution operator (Zaslavsky, 1994).

The FFPK equation is an archetype fractional equation of fractional stochastic dynamics in a (multi)-fractal phase space with fractal temporal evolution caused by the self similar and multiscale structure of islands around islands, responsible for the flights and trappins of the dynamics. The "spatial" random variable can be any physical variable, such as position in physical space, velocity in the velocity space or a dynamic field (magnetic or electric) at a certain position in physical space etc, underlying to the nonlinear chaotic dynamics. The fractional dynamics of plasma includes fractal distribution of field and currents, as well as fractal distribution of energy dissipation field.

The fractional temporal derivative $\frac{\partial^{\beta}}{\partial t^{\beta}}$ in kinetic equations allows one to take fractal-time random walks into account, as the temporal component of the strange dynamics in fractal-turbulent media. The waiting times follow the power law distribution $P(\tau) \sim \tau^{-(1+\beta)}$ since the "Levy flights" of the dynamics also follow the power law of distribution.

The asymptotics (root mean square of the displacement) of the transport process is given by $\left\langle \left| \xi \right|^{2} \right\rangle = 2Dt^{\beta/\alpha}$, while the generalized transport coefficient $D$ depends on the value of the anomalous scaling exponent $\frac{b}{a}$.

The solution of the fractal kinetic equation corresponds to Levy distributions and asymptotically to Tsallis $q$-Gaussians. According to Alemany and Zanette (1994), the set of points visited by the random walker can reveal a self-similar fractal structure produced by the extremization of Tsallis $q$-entropy. The $q$-Gaussian distribution of the fractal structure created by the strange dynamics and the extremized $q$-entropy asymptotically corresponds to the Levy distribution $P(\xi) \sim \xi^{-1-\gamma}$ where the q-exponent is related to the Levy exponent $\gamma$ by $q = \frac{3+\gamma}{1+\gamma}$. The Levy exponent $\gamma$ corresponds to the fractal structure of the points visited by the random walker.

The well known Boltzmann's formula $S = k \log W$ where $S$ is the entropy of the system and $W$ is the number of the microscopic states corresponding to a macroscopic state indicates the priority of statistics over dynamics. However, Einstein preferred to put dynamics in priority of statistics (Tarasov, 2006), using the inverse relation i.e. $W = e^{S/k}$ where entropy is not a statistical but a dynamical magnitude. Already, Boltzmann himself was using the relation $W_R = \lim_{T \to \infty} \dfrac{t_R}{T} = \dfrac{\Omega_R}{\Omega}$ where $t_R$ is the total amount of time that the system spends during the time $T$ in its phase space trajectory in the region $R$ while $\Omega_R$ is the phase space volume of region $R$ and $\Omega$ is the total volume of phase space (Tarasov, 2005). The above concepts of Boltzmann and Einstein were innovative as concerns modern $q$-extension of statistics which is internally related to the fractal extension of dynamics. According to Zaslavsky (Zaslavsky,1993; Tsallis, 2005) the fractal extension of dynamics includes simultaneously the $q$-extension of statistics as well as the fractal extension RNG theory in the Fraction at Fokker-Planck-Kolmogorov Equation (FFPK):

$$\frac{\partial^{\beta} P(x,t)}{\partial^{\beta} t} = \frac{\partial^{\alpha}}{\partial(-x)^{\alpha}}\left[ A(x) P(x,t) \right] + \frac{\partial^{\alpha+1}}{\partial(-x)^{\alpha+1}}\left[ B(x) P(x,t) \right] \qquad (108)$$

As far as the space plasma dynamics is concerned, the plasma particle and fields magnitude correspond to the $(x)$ variable in equation (108), while $P(x,t)$ describes the probability distribution of the particle-fields variables. The variables *A(x), B(x)* corresponds to the first and second moments of probability transfer and describe the wandering process in the fractal space (phase space) and time. The fractional space and time derivatives $\partial^{\beta}/\partial t^{\beta}, \partial^{\alpha}/\partial x^{\alpha}$ are caused by the multifractal (strange) topology of phase space which can be described by the anomalous phase space renormalization transform (Zaslavsky, 1993). We must notice here that the multifractal character of phase space is the mirroring of the phase space strange topology in the spatial multifractal distribution of the dynamical variables.

The $q$-statistics of Tsallis corresponds to the meta-equilibrium solutions of the FFPK equation (Tsallis, 2005; Tarasov, 2005). Also, the metaequilibrium states of FFPK equation correspond to the fixed points of Chang non-equilibrium RNG theory for space plasmas (Shlesinger, 1987; Zaslavsky, 2002) The anomalous topology of phase space dynamics includes inherently the statistics as a consequence of its multiscale and multifractal character. From this point of view the non-extensive character of thermodynamics constitute a kind of unification between statistics and dynamics. From a wider point of view the FFPK equation is a partial manifestation of a general fractal extension of dynamics. According to Tarasov (2005), the Zaslavsky's equation can be derived from a fractional generalization of the Liouville and BBGKI equations.

According also to Tarasov (2005, 2006), the fractal extension of dynamics including the dynamics of particles or fields is based on the fact that the fractal structure of matter (particles, fluids, fields) can be replaced by a fractional continuous model. In this generalization the fractional integrals can be considered as approximations of integrals on fractals. Also, the fractional derivatives are related with the development of long range correlations and localized fractal structures. In this direction the solar dynamo theory must be based at the extended fractal plasma theory including anomalous magnetic transport and diffusion, magnetic percolation and magnetic Levy random walk (Consolini, 1996). Also from a more extreme point of view, the fractal environment for the anomalous turbulence dissipation of magnetic field and plasma flows is the fractality of the space-time itself according to Shlesinger (1993). Finally we must state that the solar phase transition process corresponds to the topological phase transition process of the attracting set in the phase space of the solar dynamics.

## 5.4 Fractal topology, critical percolation and stochastic dynamics

The nonlinearity in the dynamics of the solar plasma system including fields ($\mathbf{B}$, $\mathbf{E}$) and particles can create random distributions in space of dynamical fields and material fields (bulk, velocity, pressure, temperature, fluxes, currents etc). In this section we follow Milovanov and Zimbardo (2000) and Milovanov (2012) and present some basic concepts concerning topological aspects of percolating random fields.

For any random field distribution $\psi(\mathbf{x})$ in the $n$-dimensional space ($E^n$) there exists a critical percolation threshold which divides the space $E^n$ into two topological distinct parts: Regions where $\psi(\mathbf{x}) < h_c$ marked as "empty", and regions where $\psi(\mathbf{x}) > h_c$, marked as "filled". When $\psi(\mathbf{x}) \neq h_c$, one of these parts will include an infinite connected set which is said to percolate. As the threshold $h$ changes we can find the critical threshold $h_c$ where the topological phase transition occurs, namely the nonpercolating part starts to percolate.

The geometry of the percolating set at the critical state ($h \rightarrow h_c$) is a typical fractal set for length scales between microscopic distances and percolation correlation length which diverges. The statistically self-similar geometry includes power-law behavior of the "mass" density of the fractal set such as "fractal mass density" $\sim x^{D-n}$, where $x$ is the length scale, $D$ is the Hausdorff fractal dimension which must be smaller than the dimensionality ($n$) of the embedding Euclidean space. In addition to the parameter $D$ of the fractal dimension, there is the index of connectivity $\theta$ which describes the "shape" of the fractal set and may be different for fractals even with equal values of the fractal dimension $D$. The index of connectivity $\theta$ is defined as characterizing the shortest (geodesic) line connecting two different points on the fractal set by the relation $d_\theta = (2 + \theta)/2$, where $d_\theta$ is the minimal Hausdorff dimension of the minimal (geodesic) line. The geodesic line on a self-similar fractal set ($F$) is a self affine fractal curve whose own Hausdorff fractal dimension is equal to $(2 + \theta)/2$. The index of connectivity plays an essential role in many dynamical phenomena on fractals, while it is a topological invariant of the fractal set $F$.

From the fractal dimension D and the connectivity index θ we can define a hybrid parameter $d_s = \dfrac{2D}{2+\theta}$ which is known as the spectral or the fracton dimension which represents the density of states for vibrational excitations in fractal network termed as fractions (Milovanov, 2012). The root mean square displacement of the random walker on the fractal set is given by

$$\left\langle \left|\xi\right|^2 \right\rangle \sim t^{2/2+\theta} = t^{d_\theta^{-1}} \qquad (109)$$

Where $d_\theta$ is the fractal dimension of the self-affine trajectory on the fractal set. Also, the spectral dimension which measures the probability of the random walker to return to the origin, is given by

$$P(t) \sim t^{-d_s/2} \qquad (110)$$

while the Hausdorff fractal dimension $D$ is a structural characteristic of the fractal structure $F$, the spectral dimension $d_s$ mirrors the dynamical properties such as wave excitation, diffusion etc. The fractal dimension $D$ of the fractal structure $F$ of a percolating random field distributed in the $E^n$ Euclidian space is given by $D = n - \dfrac{\beta}{v}$, where $\beta$, $v$ are the universal critical exponents of the critical percolation state (Milovanov, 2012). According to the Alexander-Orbach (AO) conjecture (Milovanov, 1997), the spectral dimension $d_s$ has been established to be equal to the value $d_s$=4/3, for all embedding dimensions $n \geq 2$. Especially, for embedding dimensions $2 \leq n \leq 5$, Milovanov (1997) has improved the AO conjecture to the value

$d_s = c \simeq 1,327$ where $c$ is the percolation constant. This constant determines the minimal fractional number of the degrees of freedom that the random walker must have to reach the infinitely remote point in the Euclidian embedding space $E^n$. According to Milovanov (1997, 2009, 2012) and Milovanov and Zelenyi (2000) the fractal topology and critical percolation theory transform the description of plasma diffusion, bulk flow and electrodynamical-MHD phenomena from classical smooth equations to fractional equations description.

The diffusion coefficient is now expressed in terms of the connectivity index $\theta = \dfrac{(\mu - \beta)}{v}$ and the Hausdorff dimension of the infinite percolation cluster $d_f = \dfrac{n - \beta}{v}$ where $\beta, \mu, v$ are the universal critical percolation parameters. The Maxwell equations for the plasma current system are transformed to fractional equations including fractional derivatives of the magnetic field given by

$$\frac{\partial^{\gamma}(\mathbf{B})}{\partial s^{\gamma}} = \frac{1}{\Gamma(1-\gamma)} \frac{\partial}{\partial s} \int_{0}^{s} \frac{\mathbf{B}(w)}{(s-w)} dw \qquad (111)$$

where $s$, $w$ are spatial variables.

The fractional extension of Maxwell equations is caused by the non-local self-similar hierarchical structuring of the plasma system, while the degree of non-locality is quantified by the connectivity index included in the power exponent $\gamma = 2/(2+\beta)$ in the singular kernel $(s-w)^{-\gamma}$. The interaction of plasma charged particles and dynamical fields described by fractional Maxwell equations and fractional transport equations causes self-consistency in the fractal distribution of dynamical (magnetic – electric) fields, of bulk plasma flow fields and energy dissipation fields. In this way, the fractal dimension and connectivity index $(\theta)$ of dynamical field's distribution is self-consistently related to the fractal dimension $(D^{+})$ and the index of connectivity $(\theta^{+})$ of material fields.

## 5.5 The solar wind self-organizing complexity

According to the experimental data analysis results and the theoretical framework of fractional dynamics solar wind plasma is a globally hierarchical, self-similar and scale invariant physical system including nonlinear and non-local internal fractional dynamics, maintaining the hierarchical structure of the turbulence. In parallel, solar wind can mirror the fractional dynamics of the solar convection zone and solar photosphere (Milovanov and Zelenyi, 1994c). Tsallis $q$-entropy principle included in his nonextensive statistical mechanics can reliably explain the solar wind self-similar hierarchical turbulent structuring and phase transition processes presented in this study. Solar wind plasma, as any other system which lives far from equilibrium, can reveal metaequilibrium stationary states NESS as critical percolation states. These nonequilibrium states, similar to Boltzmann-Gibbs thermodynamical equilibrium states, can be produced as the system tends to obtain extremization of Tsallis $q$-entropy (S$q$). The internal mechanism for this is the anomalous diffusion process in the physical space or the anomalous random walk in a hierarchical and multifractal structured phase space. The dynamics in the multifractal phase space is described by the fractional Langevin and the corresponding FFPK equations.

Moreover we conjecture that the metaequilibrium stationary states can be obtained also as the fixed points of a fractional renormalization flow equation in a fractal parameter space. This concept is the extension of the Chang's stochastic dynamics and renormalization group theory for space plasmas (Chang, 1992; Chang and Wu, 2002). Moreover, the hierarchical, self-similar, multiscale and multifractal structure of the solar wind plasma system at critical percolation and intermittent turbulent states can be obtained by the solution of the fractional MHD Langevin equation, as $N$-point correlation functions related to the functional derivative of the $q$-partition function $Z_q$ defined in the framework of non-extensive Tsallis statistical mechanics-thermodynamics (Tsallis, 2009). In the following we present a more analytical sketch of these concepts.

## 5.6 Renormalization Group (RNG) theory and space phase transition

The multifractal and multiscale intermittent turbulent character of the space plasma dynamics in the solar wind was verified in this study after the estimation of the spectrum $f(a)$ of the point wise dimensions or singularities ($a$). This justifies the application of RNG theory for the description of the scale invariance and the development of long – range correlation of the space plasma intermittent turbulence state. Generally the space plasma can be described by generalized Langevin stochastic equations of the general type:

$$\frac{\partial \varphi_i}{\partial t} = f_i (\boldsymbol{\varphi}, \mathbf{x}, t) + n_i (\mathbf{x}, t) \qquad i = 1, 2, \ldots \tag{112}$$

where $f_i$ corresponds to the deterministic process as concerns the plasma dynamical variables $\phi(\mathbf{x}, t)$ and $n_i$ to the stochastic components (fluctuations). Generally, $f_i$ are nonrandom forces corresponding to the functional derivative of the free energy functional of the system. According to Chang [Chang,1992; 1999] and Chang et al [Chang. et. al, 1978] the behavior of a nonlinear stochastic system far from equilibrium can be described by the density functional $P$, defined by

$$P (\boldsymbol{\varphi} (\mathbf{x}, t)) =$$
$$\int D (\mathbf{x}) \exp \left\{ -i \cdot \int L (\dot{\boldsymbol{\varphi}}, \boldsymbol{\varphi}, \mathbf{x}) d\mathbf{x} \right\} dt \tag{113}$$

where $L(\dot{\boldsymbol{\varphi}}, \boldsymbol{\varphi}, \mathbf{x})$ is the stochastic Lagrangian of the system, which describes the full dynamics of the stochastic system. Moreover, the far from equilibrium renormalization group theory applied to the stochastic Lagrangian $L$ generates the singular points (fixed points) in the affine space of the stochastic distributed system. At fixed points the system reveals the character of criticality, as near criticality the correlations among the fluctuations of the random dynamic field are extremely long-ranged and there exist many correlation scales. Also, close to dynamic criticality certain linear combinations of the parameters, characterizing the stochastic Lagrangian of the system, correlate with each other in the form of power laws and the stochastic system can be described by a small number of relevant parameters characterizing the truncated system of equations with low or high dimensionality.

According to these theoretical results, the stochastic space plasma system can exhibit low dimensional chaotic or high dimensional SOC like behavior, including fractal or multifractal structures with power law profiles. The power laws are connected to the near criticality phase transition process which creates spatial and temporal correlations as well as strong or weak reduction (self-organization) of the infinite dimensionality corresponding to a spatially distributed system. First and second phase transition processes can be related to discrete fixed points in the affine dynamical (Lagrangian) space of the stochastic dynamics. The SOC like behavior of plasma dynamics corresponds to the second phase transition process as a high dimensional process at the edge of chaos. The process of strong and low dimensional chaos can be related to a first order phase transition process. The probabilistic solution (113) of the generalized Langevin equations may include Gaussian or non-Gaussian processes as well as normal or anomalous diffusion processes depending upon the critical state of the system.

From this point of view, a SOC or low dimensional chaos interpretation or distinct q-statistical states with different values of the Tsallis q-triplet depends upon the type of the critical fixed (singular) point in the functional solution space of the system. When the stochastic system is externally driven or perturbed, it can be moved from a particular state of criticality to another characterized by a different fixed point and different dimensionality or scaling laws. Thus, the old SOC theory could be a special kind of critical dynamics of an externally driven stochastic system. After all SOC and low dimensional chaos can coexist in the same dynamical system as a process manifested by different kinds of fixed (critical) points in its solution space. Due to this fact, the solar convection zone dynamics may include high dimensional SOC process or low dimensional chaos or other more general dynamical process corresponding to various q-statistical states.

The non-extensive character of the magnetospheric plasma related with q-statistical metaequilibrium thermodynamics as well as the existence of long-range correlations must be harmonized with the nonlinear dynamics of the magnetospheric plasma system. The experimental results of this study press us to look for a dynamical mechanism efficient to explain the spontaneous development of the long-range correlations and reduction of the infinite degrees of freedom corresponding to the distributed character of the magnetospheric plasma system. This physical mechanism must be related with some kind of Renormalization Group Theory (RGT) adapted to the plasma dynamics (Chang, 1999) as well as to the Tsallis $q$-statistics. The basic equations of plasma physics are local, as the laws of forces and the Maxwell's equations specify the behavior of charged particles and electric and magnetic fields in the infinitesimal neighborhood of a point $\bar{x}$ in the physical space. Because of this, the derivatives in a continuum limit are the base of theoretical physics. On the other hand the nonlinear coupling of the classical fields and particles even if quantum effects can be neglected can create infinitesimal discontinues and fractalities in the field-particle phase space and non-neglected fluctuations in the infinitesimal. This character implies a fundamental statistical continuum limit even in classical deterministic systems. This is the base for the development of anomalous diffusion plasma processes, long-range correlations and scale invariance which can be amplified as the system approaches far from equilibrium dynamical critical points. Also, it is known that nonlinear dissipative dynamics with finite or infinite degrees of freedom includes the possibility of self-organizing reduction of the effective degrees of freedom and bifurcation to periodic or strange (chaotic) attractors with spontaneous development of macroscopic ordered spatiotemporal patterns. The bifurcation points of the nonlinear dynamics, corresponds to the critical points of far from equilibrium non-classical statistical mechanics and its generalizations, as well as to the fixed points of the RGT. The RGT is based in the general principle of scale invariance of the physical processes as we pass from the microscopic statistical continuum limit to the macroscopic thermodynamic limit. For introducing the mechanism of the RGT at the plasma dynamics we start with the generalized Langevin equation:

$$\dot{\Psi}_i = -\Gamma \frac{\delta F}{\delta \Psi_i} + N_i \qquad (114)$$

where $\Gamma$ is the relaxation rate and N is the noise component with the correlation:

$$\left\langle N_i(\vec{x},t) N(\vec{x}',t') \right\rangle = 2\Gamma(\vec{x}) \delta_{ij} \delta(\vec{x} - \vec{x}') \delta(t - t') \Psi_i(\vec{x},t) \qquad (115)$$

corresponding to the distributed physical magnitude of the magnetospheric plasma system. The structure of these equations (114, 115) can be produced by using the BBGKY (Bogoliubov-Born-Green-Kirkood-Yvon) hierarchy included in the Liouville equation applied at the plasma system of charged particles and fields. Also, the structure of Langevin equation ensures that the metaequilibrium distribution is always attained as $t \to \infty$. In the following we conjecture in relation with Langevin equations the generalization of the Free energy functional to the $q$-Free energy functional $F_q$ so that the Langevin equation can be related to a $q$-generalized Fokker-Planck equation of type (1). By using the q-generalized Fokker – Planck equation we can estimate the N-point correlation function ($G_N$) of the plasma system, as well as, the partition function $Z_q$ as the functional integrations:

$$G_N^q(\vec{x}_1, \vec{x}_2, ..., \vec{x}_N) = \left\langle \Psi(\vec{x}_1), \Psi(\vec{x}_2), ..., \Psi(\vec{x}_N) \right\rangle =$$
$$= \frac{1}{2} \int D[\Psi] \cdot \Phi(\vec{x}_1) \cdot \Phi(\vec{x}_2) ... \cdot \Phi(\vec{x}_N) \cdot e^{-\int F_q d^D x} \qquad (116a)$$

$$Z(J(\vec{x})) = \int D[\Psi] \cdot e^{-\int F_q(J) d^D x} \qquad (116b)$$

where $J$ is a source field and

$$Z_q = \lim_{J \to 0} Z(J(\vec{x})) \qquad (116c)$$

The $N$-point $q$-correlation function is related to the functional derivative of partition function as follows:

$$G_N^q(\vec{x}_1, \vec{x}_2, ..., \vec{x}_N) = \frac{1}{Z} \frac{\delta^N Z_q(J(\vec{x}))}{\delta J(\vec{x}_1) \cdot \delta J(\vec{x}_2) ... \cdot \delta J(\vec{x}_N)} \qquad (117)$$

For the estimation of the plasma system's partition function it makes no sense to consider fields, included in MHD or the Boltzmann-Vlassov plasma approximation which varies rapidly on scales shorter than a characteristic microscopic plasma dimension and such field should be excluded from the estimation of the partition function, as well as, from the application of scale transformation according to the RGT. This can be succeeded by a suitable cutoff parameter of the Fourier transform of the distributed plasma properties. According to Kadanoff [63] and Wilson [55] and Chang (1999) the dynamics of plasma system as it approaches the critical points includes a cooperation of all the scales from the microscopic to macroscopic level. The multiscale and holistic plasma dynamics can be produced by scale invariance principle included in the RG transformation corresponding to the flow:

$$\vec{K}^n = R_l(\vec{K}^{n-1}) = ... = R_l^n \cdot (\vec{K}^0), \ n = 0,1,2..., \qquad (118)$$

where $\vec{K}^0$ is the original parameter vector $\vec{K}^0$ and $R_l$ is the renormalization-group operator of the partition function and the density of free energy. The parameter vector $\vec{K}$ has as components the parameters $\vec{K}_1, \vec{K}_2, ..., \vec{K}_m$ the coupling constants upon which the free energy depends. The RG flow in the dynamical parameter space of vectors $\vec{K}$ is caused by a spatial change of scale as at every step of flow in the parameter space the spatial scale is rescaled according to the relation: $\vec{r}^{'} = l^{-1} \cdot \vec{r}$. As the free energy and the partition function are rescaled by the RG flow in the parameter space the correlation length ($\xi$) of the plasma fields is rescaled also according to the relation:

$$\xi\left(\vec{K}^{n}\right) = l^{-1} \cdot \xi\left(\vec{K}^{n-1}\right) = ... = l^{-n} \cdot \xi\left(\vec{K}^{0}\right) \qquad (119)$$

At the fixed points $\vec{K}^{*}$ of the RG flow, the relation:

$$\xi\left(\vec{K}^{*}\right) = l^{-1} \cdot \xi\left(\vec{K}^{*}\right) \qquad (120)$$

implies that the correlation length at the fixed point must be either zero or infinite. Also, as the zero value is without physical interest we conclude the infinite correlation of the system at the fixed point $\vec{K}^{*}$ or long-range correlation in near the fixed point. The dynamics of the system near the physical critical point corresponds to the flow of the parameter vector $\vec{K}$ at the neighborhood of the fixed point. The flow of the parameter vector $\vec{K}$ at the neighborhood of the fixed point $\vec{K}^{*}$ is a nonlinear flow in a finite dimensional space which survives the most significant physical characteristics of the original dynamics of the plasma system with infinite degrees of freedom. The representation of the infinite dimensional dynamics to finite dimensional and is possible at every instant the infinite dimensional dynamical state (state of infinite degrees of freedom) is transformed by the scale invariance vehicle to a finite dimensional dynamics in the parameter space. According to this theoretical description the solar wind plasma system can exist at district fixed points in the parameter points corresponding to the quiet and superstorm active states. Also the development of the solar wind shock corresponds to global solar dynamics phase transition process from the calm period to the shock period fixed point in the solar plasma dynamical parameter space. From the above theoretical point of view the quantative change of the non-extensive Tsallis statistics corresponds to the solar wind's system can be related to the renormalization group theory (RGT) change of the fixed point in the dynamical parameter space of the solar wind dynamics.

## Acknowledgments

This research has been co-financed by the European Union (European Social Fund – ESF) and Greek national funds through the Operational Program "Education and Lifelong Learning" of the National Strategic Reference Framework (NSRF) - Research Funding Program: Thales. Investing in knowledge society through the European Social Fund. The project is called "Hellenic National Network for Space Weather Research" coded as MIS 377274. Also, this research has been co-financed by RBRI grants # 13-02-00819 and # 12-05-00984 and Program P-22 of RAS Physical department.

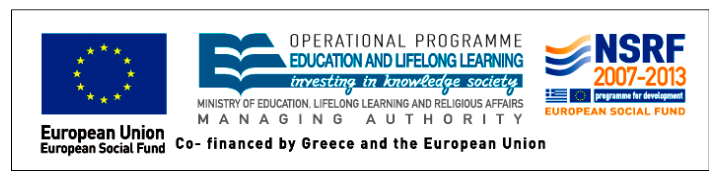

# References

Alemany, P.A. and Zanette, D.H. (1994) Fractal random walks from a variational formalism for Tsallies entropies, Phys. Rev. E, 49(2), R956-R958.

Arimitsu, T. and Arimitsu, N. (2000), Tsallis statistics and fully developed turbulence, J. Phys. A: Math. Gen., 33 L235, doi:10.1088/0305-4470/33/27/101.

Arimitsu, T. and Arimitsu, A. (2001), Analysis of turbulence by statistics based on generalized entropies, Physica A, 295, 177 – 194.

Arimitsu, T. and Arimitsu, N., (2002) Analysis of velocity fluctuation in turbulence based on generalized statistics, J. Phys.: Condens. Matter, 14, 2237–2246.

Arneodo, A., Bacry, E., Muzy, J.F. (1995) The thermodynamics of fractals revisited with wavelets, Physica A 213, 232-275.

Arneodo, A. et al. (1997) Oscillating Singularities on Cantor Sets: A Grand-Canonical Multifractal Formalism, J. Statist. Phys. 87(1/2), 179-209.

Baldovin, F. and Stella, A. L. (2007) Central limit theorem for anomalous scaling due to correlations, Phys. Rev. E 75, 020101 (R).

Bruno, R. et al. (2001) Identifying intermittency events in the solar wind, Planetary and Space Science, 49 (12), 1201–1210.

Burlaga, L.F., Ness, N.F. and McDonald F.B. (1987), Large-scale fluctuations between 13 AU and 25 AU and their effects on cosmic rays, J. Geophys. Res., 92(A12), 13647–13652, doi:10.1029/JA092iA12p13647.

Burlaga, L. (1991a) Multifractal structure of the interplanetary magnetic field: Voyager 2 observations near 25AU, 1987–1988, Geophys. Res. Lett., 18, pp. 69–72

Burlaga, L.F. (1991b) Intermittent turbulence in the solar wind. J. Geophys. Res., 96 (A4), pp. 5847–5851, doi:10.1029/91JA00087.

Burlaga, L. (1991c) Multifractal structure of speed fluctuations in recurrent streams at 1AU and near 6AU, Geophys. Res. Lett., 18, pp. 1651–1654

Burlaga, L. F. (1993), Intermittent turbulence in large-scale velocity fluctuations at 1 AU near solar maximum, J. Geophys. Res., 98(A10), 17467–17473, doi:10.1029/93JA01630.

Burlaga, L. F., and M. A. Forman, M.A. (2002) Large-scale speed fluctuations at 1 AU on scales from 1 hour to ≈1 year: 1999 and 1995, J. Geophys. Res., 107(A11), 1403, doi:10.1029/2002JA009271.

Burlaga, L. F. et al., (2003), Evolution of the multiscale statistical properties of corotating streams from 1 to 95 AU, J. Geophys. Res., 108, 1305, doi:10.1029/2003JA009841, A7.

Burlaga, L.F. and Vinas, A.F., (2005) Triangle for the entropic index q of non-extensive statistical mechanics observed by Voyager 1 in the distant heliosphere, Physica A, 356, 375-384.

Burlaga L.F. et al. (2007) Tsallis distributions of magnetic field strength variations in the heliosphere: 5 to 90 AU, JGR, 112, A07206, doi:10.1029/2006JA012213.
Buti, B. (1996) Chaos and turbulence in solar wind, Astrophysics and Space Science, 243(1), pp 33-41.

Carbone, V. (1993) Cascade model for intermittency in fully developed magnetohydrodynamic turbulence, Physical Review Letters, 71(10), p. 1546-1548.

Carbone, V. (1994) Scale similarity of the velocity structure functions in fully developed magnetohydrodynamic turbulence, Phys. Rev. E 50 (2), R671–R674.

Carbone V., Veltri P., Bruno R. (1995) Experimental evidence for differences in the extended self-similarity scaling laws between fluid and magnetohydrodynamic turbulent flows, Phys. Rev. Lett., 75, pp. 3110–3120.

Carbone, V., Veltri, P. and Bruno, R. (1996) Solar wind low-frequency magneto hydrodynamic turbulence: extended self-similarity and scaling laws, Nonlinear Process in Geophysics, 3, 247-261.

Chame A. and de Mello E.V.L., (1994) The fluctuation-dissipation theorem in the framework of the Tsallis statistics, Journal of Physics A: Mathematical and General, Volume 27, Issue 11, pp. 3663-3670.

Chang, T. (1992) "Low-Dimensional Behavior and Symmetry Braking of Stochastic Systems near Criticality Can these Effects be observed in Space and in the Laboratory", IEEE, 20(6), p. 691-694.

Chang T. (1999) Self-organized criticality, multi-fractal spectra, sporadic localized reconnections and intermittent turbulence in the magnetotail, Phys.Plasmas 6 (11), 4137–4145.

Chang, T. et al. (1978), A closed-form differential renormalization group generator for critical dynamics, Phys. Lett. 67A(4), 287-290.

Chang, T., VVedensky, D.D. and Nicoll, J.F., (1992) Differential Renormalization-Group Generators for Static and Dynamic Critical Phenomena, Physics Reports, 217(6), 279-360.

Chang, T. and Wu, C. (2002) "Complexity" and anomalous transport in space plasmas, Physics of Plasmas, 9(9), 3679-3684.

Chang T. *et al.,* (2003) Complexity Forced and/or Self Organized Criticality, and topological phase transitions in space plasmas, Space Science Reviews 107, 425-445.

Consolini, G. et al., (1996) Multifractal Structure of Auroral Electrojet Index Data, Phys. Rev. Lett. 76, 4082–4085.

Consolini, G. and Chang, T., (2001) Magnetic Field Topology and Criticality in Geotail Dynamics: Relevance to Substorm Phenomena, Space Sc. Rev., 95, 309-321.

Consolini, G., et al. (2005) Complexity and topological disorder in the earth's magnetotail dynamics in Nonequilibrium Phenomena in Plasmas, edited by Sharma A.S. and Kaw P.K., pp. 51–70, Springer.

Ferri, G.L. , Reynoso Savio, M.F., Plastino A. (2010) "Tsallis q-triplet and the ozone layer", Physica A, 389, p. 1829-1833.

Fraser, A.M. and H.L.Swinney, (1986) Independent coordinates for strange attractors from mutual information, Phys. Rev. A 33, 1134–1140.

Frisch U., (1996), Turbulence, Cambridge University Press, UK, pp. 310. ISBN 0521457130.
Gonzales, J.L. et al. (2011) Nonadditive Tsallis entropy applied to the Earth's climate Physica A 390, 587-594.

Horbury, T.A., Balogh, A. (1997) Structure function measurements of the intermittent turbulent cascade Nonlinear Process. Geophys., 4 , pp. 185–199.

Leitner, M., Vörös, Z. and Leubner, M.P. (2009), Introducing log-kappa distributions for solar wind analysis, J. Geophys. Res., 114, A12104,doi:10.1029/2009JA014476.

Macek, W.M., (2006) Modeling multifractality of the solar wind, Space Science Reviews, 122: 329–337, DOI: 10.1007/s11214-006-8185-z

Macek, W. (2007) Multifractality and intermittency in the solar wind, Nonlin. Processes Geophys., 14, 695–700.

Macek, W. et al. (2012) Observation of the multifractal spectrum in the heliosphere and the heliosheath by Voyager 1 and 2, JGR, 117, A12101, doi:10.1029/2012JA018129.

Mainardi, F. (2010) Fractional calculus and waves in linear viscoelasticity, Imperial College Press, Signapore.

Marsch, E., Liu, S. (1993) Structure functions and intermittency of velocity fluctuations in the inner solar wind, Ann. Geophys., 11, pp. 227–238.

Marsch E. and Tu CY, (1997), Intermittency, non-Gaussian statistics and fractal scaling of MHD fluctuations in the solar wind, Nonlinear Processes in Geophysics, 4(2),101-124.

Meneveau, C. and Sreenivasan, K.R. (1987) Simple multifractal cascade model for fully developed turbulence, Phys. Rev. Lett. 59(13), 1424-1427.

Milovanov, A.V. (1997) Topological proof for the Alexander-Orbach conjecture, Phys. Rev. E 56(3), 2437-2446.

Milovanov, A.V., (2001) Stochastic dynamics from the fractional Fokker-Planck-Kolmogorov equation: Large- scale behavior of the turbulent transport coefficient, Phys. Rev. E 63, 047301.

Milovanov, A.V. (2009) Pseudochaos and low-frequency percolation scaling for turbulent diffusion in magnetized plasma, Phys. Rev. E 79(4), 046403.

Milovanov, A.V. (2012) Percolation models of self organized critical phenomena, arXiv: 207.5389.

Milovanov, A.V. and Zelenyi, L.M. (1993) Applications of fractal geometry to dynamical evolution of sunspots, Phys. Fluids B **5**, 2609, http://dx.doi.org/10.1063/1.860698.

Milovanov, A.V. and Zelenyi, L.M. (1994a) Development of Fractal Structure in the Solar Wind and Distribution of Magnetic Field in the Photosphere, in Solar System Plasmas in Space and Time (eds J. L. Burch and J. H. Waite), American Geophysical Union, Washington, D. C.. doi: 10.1029/GM084p0043.

Milovanov, A.V. and Zelenyi, L.M. (1994b) Fractal cluster in the solar wind Advances in Space Research, 14 (7), 123–133.

Milovanov, A.V. and Zelenyi, L.M. (1999) Fracton Excitations as a Driving Mechanism for the Self-Organized Dynamical Structuring in the Solar Wind, Astrophysics and Space Science, 264(1-4), 317-345.
Milovanov, A.V. and Zelenyi, L.M., (2000) Functional background of the Tsallis entropy: "coarse-grained" systems and "kappa" distribution functions, Nonlinear Processes in Geophysics, 7, 211-221.

Milovanov A.V. and Zelenyi L.M. (2001) ''Strange'' Fermi processes and power-law nonthermal tails from a self-consistent fractional kinetic equation, Phys. Rev. E, 64, 052101.

Milovanov, A.V. and Zimbardo, G. (2000) Percolation in sign-symmetric random fields: Topological aspects and numerical modeling, Phys. Rev. E 62(1), 250-260.

Pavlos et al. (1991) Chaotic Dynamics in Astrophysics and space physics, 1st general conference – Balkan Physical Union

Pavlos, G.P., et al. (1992a) Evidence for chaotic dynamics in the outer solar plasma and the earth magnetosphere, Chaotic Dynamics: Theory and Practice, NATO ASI Series B: Physics, 298, 327-339.

Pavlos, G.P., et al. (1992b) Evidence for strange attractor structures in space plasmas, *Annales Geophysicae*, 10(5), 309-322.

Pavlos et all. (2013) Universality of Tsallis Non - Extensive Statistics and Time Series Analysis: Theory and Applications, Submitting for publication in Physica A.
Riazantseva, M.O. et al. (2005) Sharp boundaries of small- and middle-scale solar wind structures. J. Geophys. Res., v.110, A12110.

Riazantseva, M.O. and Zastenker, G.N. (2008) Intermittency of Solar Wind Density Fluctuations and Its Relation to Sharp Density Changes, *Cosmic Research*, 46(1), 1–7.

Riazantseva, M.O. et al. (2010) Intermittency Of Solar Wind Ion Flux And Magnetic Field Fluctuations In The Wide Frequency Region From 10-5 Up To 1 Hz And The Influence Of Sudden Changes Of Ion Flux in Twelfth International Solar Wind Conference, ed. Maksimovic M.,Issautier K., Meyer-Vernet N., Moncuquet M. and Pantellini F., AIP.

Ruzmaikin, A. et al. (1995) Intermittent turbulence in solar wind from the south polar hold, J. Geophys. Res., 100, 3395–3404.

Safrankova, J. et.al. (2013) Fast Solar Wind Monitor (BMSW): Description and First Results, Space Science Reviews, 175 (1-4), 165-182.

Shlesinger, M.F. et al. (1987) Lévy dynamics of enhanced diffusion: Application to turbulence, Phys. Rev. Lett., 58, 1100–1104.

Shlesinger, M.F (1988) Fractal time in condensed matter, Annual Review of Physical Chemistry 39 (1), 269-290.

Shlesinger, M.F., Zaslavsky, G.M. and Klafter, J. (1993) Strange kinetics, Nature, 363, 31.

Sorriso, V. et al., (1999) Intermittency in the solar wind turbulence through probability distribution functions of fluctuations, Geophysical Research Letter, 26 (13), 1801–1804.

Strumik, M. and Macek, W. M. (2008) Testing for Markovian character and modeling of intermittency in solar wind turbulence, Phys. Rev. E, 78, 026414.

Tarasov, V.E. (2005) Fractional Fokker-Planck Equation for Fractal Media, Chaos, 15 (2), 023102.

Tarasov, V.E. (2006) Magnetohydrodynamics for fractal media, Phys. Plasmas 13, 052107.

Theiler, J. (1990) Estimating fractal dimension, JOSA A, 7(6), 1055-1073.

Tsallis, C., (1988), Possible generalization of Boltzmann-Gibbs statistics, Journal of Statistical Physics, 52,1-2, 479-487.

Tsallis, C. (2002), "Entropic nonextensivity a possible measure of complexity" Chaos, Solitons and Fractals, 13, 371-391.

Tsallis, C. (2004a) Dynamical scenario for nonextensive statistical mechanics, Physica A, 340, 1 – 10.

Tsallis C. (2004b), Nonextensive statistical mechanics: construction and physical interpretation, Nonextensive entropy – interdisciplinary applications Ed. Murray GM and Tsallis C., Oxford Univ. Press, 1-53.

Tsallis, C. (2004c) What should a statistical mechanics satisfy to reflect nature?, Physica D 193, 3-34.

Tsallis, C. (2009), Introduction to Non-extensive Statistical Mechanics, Springer.
Tu, C.-Y., Marsch, E., Rosenbauer, H. (1996) An extended structure function model and its application to the analysis of solar wind intermittency properties, Ann. Geophys., 14, 270–285.

Umarov, S. et al. (2008) On a $q$-Central Limit Theorem Consistent with Nonextensive Statistical Mechanics, Milan j. math. 76 , 307–328.

 Zaslavsky G.M., Renormalization group theory of anomalous transport in systems with Hamiltonian chaos, Chaos 4 (1) (1993) 25.

Zaslavsky, G.M. (2002) Chaos, fractional kinetics, and anomalous transport, Physics Reports, 371, 461–580.

Zastenker, G.N. et. al. (2013) Fast measurements of parameters of the Solar Wind using the BMSW instrument, Cosmic Research, 51 (2), 78-89.

Zelenyi, L.M. and Milovanov, A.V. (1991) Fractal Properties of Sunspots, Soviet Astronomy Letters, 17(6), 425.

L.M. Zelenyi, A.V. Milovanov, Applications of Lie groups to the equilibrium theory of cylindrically symmetric magnetic flux tubes, Sov. Astron. Lett. 36 (1) (1993) 74.

Zelenyi, L. M. and Milovanov, A. V. (2004) Fractal topology and strange kinetics: from percolation theory to problems in cosmic electrodynamics, Pysics-Uspekhi, 47(8), 749-788.